\documentclass[pra,a4paper,nofootinbib,superscriptaddress,aps]{revtex4-2}
\usepackage{hyperref}
\usepackage{graphicx}
\usepackage{mathtools}
\usepackage{amssymb}
\usepackage{amsthm,dsfont}
\usepackage{amsmath}
\usepackage{color}
\allowdisplaybreaks
\theoremstyle{plain}

\usepackage[normalem]{ulem}
\usepackage{chngpage}
\usepackage{appendix}
\usepackage{fancyhdr}
\usepackage[caption=false]{subfig}
\useunder{\uline}{\ul}{}

\newcommand{\tr}{\text{Tr}}
\newcommand{\abs}[1]{\left|#1\right|}

\newcommand{\bra}{\left \langle}
\newcommand{\ket}{\right \rangle}
\newcommand{\bq}{\begin{equation}}
\newcommand{\eq}{\end{equation}}
\newcommand{\ba}{\begin{align}}

\definecolor{red}{rgb}{0.9,0,0}
\definecolor{green}{rgb}{0,0.8,0}
\definecolor{blue}{rgb}{0,0,0.8}
\definecolor{cautionred}{rgb}{1.0,0,0}

\definecolor{maroon}{rgb}{0.7,0,0}

\definecolor{ngreen}{rgb}{0.3,0.7,0.3}

\definecolor{golden}{rgb}{0.8,0.6,0.1}

\begin{document}
\raggedbottom
\bibliographystyle{unsrt}
\title{Testing Generalised Uncertainty Principles through Quantum Noise}
\author{Parth Girdhar}
\affiliation{Centre for Engineered Quantum Systems, School of Physics, The University of Sydney, Sydney, NSW 2006, Australia}
\author{Andrew C. Doherty}
\affiliation{Centre for Engineered Quantum Systems, School of Physics, The University of Sydney, Sydney, NSW 2006, Australia}

\begin{abstract}
 Motivated by several approaches to quantum gravity, there is a considerable literature on generalised uncertainty principles particularly through modification of the canonical position-momentum commutation relations. Some of these modified relations are also consistent with general principles that may be supposed of any physical theory. Such modified commutators have significant observable consequences. Here we study the noisy behaviour of an optomechanical system assuming a certain commonly studied modified commutator. From recent observations of radiation pressure noise in tabletop optomechanical experiments as well as the position noise spectrum of Advanced LIGO we derive bounds on the modified commutator. We find how such experiments can be adjusted to provide significant improvements in such bounds, potentially surpassing those from sub-atomic measurements. 
\end{abstract}
\maketitle

\tableofcontents
\newpage

\section{Introduction}

Quantum mechanics and the general theory of relativity have both been successful at explaining most of our observations of the universe, the former is an accurate framework for predictions of phenomena at the microscopic scale while the latter explains gravity at the macroscopic scale. However quantizing gravity in the standard way has well-known difficulties due to the non-renormalizability of the gravitational interaction \cite{weinberg1995quantum}. Furthermore the combined strong quantum and gravitational effects expected at the Planck scale, and also around black hole horizons, have brought the postulates of these theories into question\cite{kiefer2006quantum}\cite{Almheiri2013} \cite{marolf2016violations}. 

One attempt to resolve this dilemma is to introduce the concept of a `minimal length' of space, challenging the paradigm from relativity that space time is continuous \cite{garay1995quantum} \cite{hossenfelder2013minimal}. According to quantum mechanics the uncertainty principle $\Delta x \Delta p\geq \frac{\hbar}{2}$ provides no restriction on measuring either the position or momentum of a particle with arbitrary precision (where precision is based on the spread of outcomes over many repeated measurements), though not both. But if there is an effective minimal length then it may mean that there is a fundamental limit to the precision of a position measurement. This can be accounted for by adding terms dependent on the variance of momentum on the right hand side of the standard uncertainty principle. Such generalised uncertainty principles (GUP's) have been argued to exist in string theory\cite{witten1996reflections}\cite{gross1988string} and the finite bandwidth approach to quantum gravity\cite{PhysRevLett.103.231301}\cite{PhysRevD.86.085017} but also follow from thought experiments involving quantum physics applied to black holes\cite{scardigli1999generalized}, relative locality\cite{amelino2011principle} and double special relativity\cite{amelino2002doubly}.

But how does one find evidence for minimal length? Or rule out specific possibilities, such as modified commutators, based on experimental data? On top of the theoretical difficulties in quantising gravity it is generally thought that experiments that can probe quantum gravity effects directly require energies much higher than currently available. Most effort has been devoted to find indirect signatures of quantum gravity in high energy particle collisions or astrophysics e.g. supersymmetry \cite{haber1985search}\cite{martin2010supersymmetry} and modified gamma ray burst dispersion relations \cite{amelino1998tests}. Though no direct evidence for quantum gravity has been found, constraints on modifications of the standard position-momentum commutation relation (equivalent to GUP's), have been derived from a variety of sources and offer a novel route to test quantum gravity since they may be probed via precision measurements in low-energy experiments. Such modified relations are also motivated by the fact that they generalise quantum mechanics without compromising certain physical principles. Other theories like extra space dimensions, axionic dark matter, dark energy models and semi-classical gravity, can also been constrained by table top experiments involving torsion balances\cite{adelberger2013torsion}, induced optomechanical interactions\cite{PhysRevD.93.124049}\cite{yang2013macroscopic}\cite{rider2016search} and atom interferometry\cite{arvanitaki2018search}\cite{hamilton2015atom}.

The possibility of using optomechanics to constrain GUP's was considered in \cite{pikovski2012probing}. There they showed that controlled interactions of optical pulses with a mechanical oscillator could lead to dependence of the output optical field on the mechanical canonical commutation relation. It was suggested that the ability to measure the average value of this field to high precision can lead to significant improvements in probing the GUP. Other proposals to probe GUP's in the dynamics of oscillators were considered in \cite{bawaj2015probing} and \cite{bushev2019testing}.

In this spirit, in this paper we investigate how modifying the standard commutation relation, which gives rise to a GUP, affects macroscopic oscillator motion in a noise bath environment that is typical of sensitive tabletop and interferometric experiments.  We find to first order the explicit modified noise spectrum associated with Brownian motion and quantum radiation pressure noise if the oscillator is driven by an optical source. This results in constraints on the GUP via the overall noise spectrum observed in Advanced LIGO (aLIGO) and recent experiments that have reported observation of quantum radiation pressure noise on mechanical membranes. We find that current constraints from the spectra close to oscillator resonance frequency or at frequencies in a free-mass limit are comparable to the best available to date. These and related experimental scenarios are optimised to show that by adjusting optomechanical parameters by a few orders of magnitude or with additional external driving of the high quality factor oscillator new bounds are feasible.

\section{Background}

The uncertainty principle can be generalised in several ways to incorporate minimal length. If this is obtained from the Robertson uncertainty relation $\Delta x \Delta p\geq \frac{1}{2}|\bra\left[x,p\right]\ket|$ then the existence of a minimal length can be loosely translated into the question: is the canonical commutation relation modified? 

Note that the existence of an absolute minimal length is in apparent conflict with Lorentz invariance. This poses very serious challenges to a consistent theory of quantum gravity, see for example the discussion here~\cite{collins2004lorentz}. There are also very strong experimental bounds on violations of Lorentz invariance in many physical systems, see for example~\cite{bluhm2006overview, liberati2013tests, tasson2014we}. In this paper we are just interested in how strongly we can hope to bound the modified commutators that arise in these theories. However, see \cite{kempf2018quantum} for a recent discussion on reconciling minimal length with relativity via covariant band-limitation.

Several modified commutation relations have been considered in the literature. For example, the simplest modified isotropic (direction symmetric) translational invariant commutator is \cite{kempf1995hilbert}:

\bq\label{eq:beta}
\left [x,p\right]=i\hbar\left(1+\beta_0\left(\frac{p}{M_Pc}\right)^2\right)=i\hbar\left(1+Ap^2\right)\\
\eq

where $\hbar=h/2\pi$, $c$ is the speed of light, $M_{P}$ is the Planck mass and $A=\beta_0/{M_P^2c^2}$.

A closely related uncertainty principle has been discussed as a caricature of the situation in string theory \cite{witten1996reflections}:

\ba
\Delta x \Delta p \geq \frac{\hbar}{2}\left(1+\frac{\beta_{0}}{\left(M_{P}c\right)^2}\Delta p^2\right).
\end{align}

This inequality implies that $\Delta x\geq L_{P}\sqrt{\beta_{0}}$. So if $\beta_{0}=1$ then the minimal length is the Planck length. This GUP is closely related to the modified commutator model (\ref{eq:beta}), at least when $\langle p\rangle =0$. Reference \cite{witten1996reflections} gives a qualitative explanation of how this arises from T-duality and a consideration of the expected behaviour of a Heisenberg microscope in the framework of string theory which, in short, implies that an effective minimal length emerges from the equivalence of high energy (small radius) and low energy (large radius) string physics. This uncertainty principle may also be arrived via a theory involving intrinsic uncertainty of spatial translations\cite{milburn2006lorentz}.

Maggiore \cite{maggiore1993algebraic} considered the problem of deriving the most general commutation relations provided that the angular momenta satisfy usual SU(2) commutation relations and satisfy usual commutation relations with the position and momenta operators, momenta commute between themselves so the translation group is not deformed, and the commutation relations depend on a parameter with dimensions of mass so that in a limit the standard relations result.  

The one-dimensional canonical commutator then has a unique form:

\bq\label{mumodified}
\left[x,p\right]=i\hbar\sqrt{1+2\mu_0\frac{\left(\frac{p}{c}\right)^2+m^2}{M_P^2}}
\eq

where $\mu_0$ is a free numerical parameter, m is the mass of the particle. This shows that given some well-motivated constraints a consistent modification of quantum mechanics of this type is possible. 
In the limit of sufficiently small $\mu_0$, mass and momentum this model for a modified commutator reduces to the previous one.

Another model for modified commutators arose in an approach to quantum gravity known as `doubly special relativity' theory \cite{amelino2002doubly}. There it is proposed that the Lorentz transformations are modified in order to keep an invariant energy-momentum scale and maximal possible momentum. In \cite{ali2009discreteness} the following commutator was given consistent with doubly special relativity:

\bq\label{deltamodified}
\left [x,p\right]=i\hbar\left(1-\delta_0\frac{p}{M_Pc}+\delta_0^2\left(\frac{p}{M_Pc}\right)^2\right)=i\hbar\left(1-Cp+C^2p^2\right)\\
\eq

where $\delta_0$ is a numerical parameter that quantifies interaction, $C=\delta_0/{M_Pc}$. Qualitatively similar commutation relations that imply a maximum momentum together with minimal length were studied in \cite{pedram2012higher} and \cite{shababi2017two}.

In this work we will focus mainly on the $\beta$-modified commutator or (\ref{eq:beta}), although our techniques are relevant to all these examples. This is to streamline the discussion and because bounds on $\beta_0$ are typically much weaker than in the other scenarios.

Constraints on $\beta_0$ have been inferred from a variety of experimental observations. Some of the best constraints on $\beta_{0}$ discussed in the literature are shown in table \ref{table:table1}. The general reason for the large order of magnitudes of the upper bounds is that the modifications are applied in these experiments to particles with momenta much smaller than the Planck momentum $M_{P}c$ that appears in the right hand side of each commutator and the parameters being probed constrain the size of the minimal length in Planck length units.

\begin{table}[h]

          \caption{Experimental bounds on $\beta_{0}$}

          \label{table:table1} 

          \begin{center}

                   \begin{tabular}{ | l | l |p{5cm} |}

                             \hline

                             Experiment & Max $\beta_{0}$\\ \hline

                             Lamb shift \cite{das2008universality}&$10^{36}$ \\ 

                             High energy particle collisions \cite{das2008universality} & $10^{32}$ \\

                             AURIGA detector \cite{marin2014investigation} & $10^{33} $ \\

                             Scanning tunnelling microscope \cite{das2008universality,pikovski2012probing} & $10^{33}$\\

                             Harmonic oscillators (frequency shifts) \cite{bawaj2015probing,bushev2019testing}  & $10^{6} - 10^{20}$\\

                             \hline

                   \end{tabular}

          \end{center}

\end{table}

One important feature of these various bounds on $\beta_0$ is whether the experiments concern the modification of the commutators of macroscopic systems, or of elementary particles. The standard canonical commutator $\left[x,p\right]= i \hbar$ is component-wise scale invariant in the sense that if it applies to components of an object then it also applies to the object's centre of mass. However this is not true in general for modified commutators \cite{pikovski2012probing,amelino2013challenge}. If we assume that the modified commutator applies to elementary particles then it is necessary to make some further assumptions about the model in order to determine how the commutation relations of a composite system should be determined. The relationship between the modification parameter e.g. $\beta_{0}$ for the centre of mass of $N$ identical systems is derived in \cite{quesne2010composite} to have approximately $1/N^{2}$ scaling with respect to the corresponding \textit{effective} modification parameter ${\beta_{c}}$ for a single system obeying \eqref{eq:beta}, up to a term involving the relative momenta of the systems. If this scaling is applied, bounds on $\beta$ for individual systems are greatly weakened. For example, this was applied to significantly weaken a constraint on $\beta$ reported in \cite{PhysRevD.66.026003} from analysis of planetary motion. The constraints in \cite{marin2014investigation}, \cite{bawaj2015probing} and \cite{bushev2019testing} also apply to macroscopic systems and are thus also weakened by this component-dependent factor. However in table \ref{table:table1} the bounds from measurements of Lamb Shift, high energy particle collisions (we use the updated value of 13 TeV as in the Large Hadron Collider) as well as electron tunnelling directly refer to elementary particles. Even some heuristic calculations of $\beta_e$ from gravitational phenomena have been attempted, given values of order 1 \cite{scardigli2017gup,luciano2019gup}. The possibility of other scaling behaviour of the bounds as a function of number of components was recently studied \cite{kumar2019quantum}. As we discuss in more detail in the following we will report bounds on both $\beta_0$ and on $\beta_e$, the value of $\beta$ applying to the elementary particles that make up a macroscopic oscillator.

As mentioned earlier, \cite{pikovski2012probing} studied how quantum optical control and readout of a mechanical oscillator, which would have a relatively large momentum compared to experiments on subatomic particles, can be used to significantly improve the constraints on these parameters. 
They show that the challenge is to perform a large number of measurement runs in which each run involves measuring a large photon number coherent state interacting with an optically cooled macro-oscillator in a cavity of high finesse, but in principle if the commutators are applied to the centre of mass the parameters can be constrained to the Planck scale i.e. $\beta_{0},\mu_{0}, \delta_{0} \sim1$.

In recent experiments it was shown how other techniques probing the mean or variance of the position of harmonic oscillators lead to new bounds on the parameters. In \cite{bawaj2015probing} and more recently \cite{bushev2019testing} the commutator \eqref{eq:beta} was associated with perturbation of the resonance frequency of a harmonic oscillator by an amount dependent on the initial amplitude of motion (`amplitude-frequency effect'). Also the existence of third-harmonics was discussed in \cite{bawaj2015probing}. From these techniques the bounds at the bottom of table \ref{table:table1} were inferred.

In this work we similarly explore the potential of macroscopic oscillators to constrain the parameters in these modified commutator theories with a focus on the modified quantum noise that an oscillator would experience due to its interaction with an environment. The simplest scenario is interaction with a thermal bath but by adding a radiation field an optomechanical interaction arises and so the oscillator experiences radiation pressure noise (on top of a mean classical radiation pressure)\cite{aspelmeyer2014cavity}. The cavity light field can be used to read out the position of the oscillator and the noise on this signal could potentially provide bounds on $\beta_0$. This is a promising route to probe the canonical commutator as highly sensitive position measurements of mechanical oscillators can be performed experimentally for a very wide range of masses. For example, radiation pressure noise has recently been observed on an oscillator \cite{purdy2013observation} and is expected to have a significant presence in LIGO interferometers in the near future \cite{martynov2016sensitivity}. 

A somewhat related earlier idea from \cite{amelino1999gravity} and \cite{ng2000measuring} was to exploit the sensitivity of gravitational wave interferometers to probe heuristic models of quantum foam, see also \cite{adler2000detectability}\cite{ng2000wigner}. Also an experiment has been conducted attempting to find position variation correlations, associated with other theories of quantum gravity, between two  interferometers\cite{chou2017holometer}. A closely related idea was discussed in \cite{bosso2018potential} where the commutation relations of canonical operators of optical fields were modified to infer perturbed radiation pressure and shot noise in interferometers. Strong bounds on the commutators could be derived in this case. In this paper we will only focus on the modification of commutation relations of massive particles and not on modified theories of electrodynamics. So in our setup only the commutators of the mechanical oscillator are modified.

\section{Modified Noise Spectrum}

In this section we use a perturbation expansion to derive the mechanical noise spectrum for an oscillator driven by thermal and radiation pressure noise with the modified canonical commutator \eqref{eq:beta}. We proceed by an analysis of the quantum Langevin equations for this system.

\subsection{Setup}
We consider a typical optomechanical system, as discussed in \cite{aspelmeyer2014cavity} for example. The system is composed of an optical field interacting with a mechanical oscillator inside a cavity and can be associated with the Hamiltonian:  
\bq
H=\hbar \omega_{c} a^{\dagger} a+\frac{1}{2}m\Omega^2 x^2-\hbar G a^\dagger a x+\frac{p^2}{2m}
\label{eq:optomechH}
\eq
Here $a, a^{\dagger}$ are creation and annihilation operators for the optical field, and $x, p$ are the position and momentum operators for the mechanical oscillator. $\omega_{c}$ is the optical cavity angular frequency,  $m$ is the mass of the oscillator, $\Omega$ is its resonance angular frequency, and $G=\omega_{c}/L$ is a optomechanical coupling constant between oscillator and field. In this model we assume that only the mechanical commutators are modified, for the optical commutators we have $[a,a^\dagger]=1$ as usual.

Let the canonical commutation relation be that in equation \eqref{eq:beta}. We wish to repeat a typical analysis of this optomechanical system in the limit that $A$ is small. While there are various approaches to this, we will proceed by rewriting the Hamiltonian (\ref{eq:optomechH}) in terms of a modified momentum operator $\tilde{p}$ that satisfies the usual commutation relations $[x,\tilde{p}]=i\hbar$. This will allow us to perform the calculation using standard perturbation theory. Since it is anticipated that the modification to the commutator is very small we will calculate only to first order in $A$.

The modified momentum operator $\tilde{p}$ is defined by the following equation
\begin{equation}
\tilde{p}=p - Ap^3/3 .
\end{equation}
It is easy to see that $[x,\tilde{p}]=i\hbar$ up to first order in A \cite{quesne2010composite}. 
Rewriting the Hamiltonian (\ref{eq:optomechH}) in terms of $\tilde{p}$ and retaining terms up to first order in A we find
\ba\label{EffectiveHamiltonian}
H\simeq\hbar \omega_{c} a^{\dagger} a+\frac{1}{2}m\Omega^2 x^2-\hbar G a^\dagger a x+\frac{\tilde{p}^2}{2m}+\frac{A\tilde{p}^4}{3m}
\end{align}
Thus the modification of the momentum operator effectively results in a potential term $V=A\tilde{p}^{4}/3m$. 
This bears similarity with the Duffing oscillator, which has a potential quartic in position, and assuming the dynamical equations of quantum mechanics are unaffected in a theory with a GUP it implies the equation of motion will be non-linear in position. We will assume that the mechanical oscillator is also coupled to a thermal bath with damping rate $\gamma$ and temperature $T$.

We adopt the standard approach of \cite{aspelmeyer2014cavity}, linearising the optical field around the mean field $\alpha$: $a=\alpha+\delta a$. $\alpha$ describes the steady-state field amplitude in the cavity and depends on the drive power $P$ and the optical decay rate $\kappa$. We will assume throughout that the cavity is driven on resonance. Without loss of generality we can choose $\alpha$ to be real. We keep only terms linear in $\delta a, \left(\delta a\right)^\dagger$. Furthermore we apply linear perturbation theory in $A$ to position and momentum: $x\simeq x_{0}+\delta{x}$ where $x_0$ is the solution for $x$ from the Hamiltonian with standard commutator $\left(A=0\right)$ and $\delta x$ is the perturbation to first-order in $A$, and similarly $\tilde{p}\simeq p_{0}+\delta p$. Writing the Langevin equations for the mechanics alone we have
\ba
\dot{x}&\simeq \dot{\left(x_{0}+\delta x\right)}=\left(p_{0}+\delta p\right)/m+4Ap_{0}^3/3m\\
\dot{\tilde{p}}&\simeq \dot{\left(p_{0}+\delta p\right)}=-m\Omega^2 \left(x_{0}+\delta x\right)-\gamma m\dot{\left(x_{0}+\delta x\right)}+f
\end{align}
The total driving force is $f=f_T+f_D$ where $f_T$ is the stochastic thermal force and $f_D$ describes the other forces on the oscillator, including the fluctuations in the optomechanical force $ \hbar G \alpha \left(\delta a + \delta a^\dagger\right)$.

Note that the form of the damping term $\gamma m\dot{\left(x_{0}+\delta x\right)}$, called `viscous damping', follows from the usual derivation as in \cite{gardiner2004quantum} that involves Heisenberg equations of motion under a Hamiltonian that contains kinetic and potential terms for the system oscillator and a thermal bath of harmonic oscillators weakly coupled to the system, provided that the modification to commutation relations of bath oscillators has negligible influence compared to that on the system. Unlike the standard case this term is inequivalent to $\gamma p$. Another common model of damping for mechanical modes, known as `structural damping', can be handled heuristically by making $\gamma$ frequency-dependent, see \ref{ligomodelling} for details.

Thus for $x_{0}, p_{0}$ we obtain the system of equations:
\ba\label{eq:system1}
\begin{pmatrix}
\dot{x_{0}}\\ 
\dot{p_{0}}
\end{pmatrix}
=
\begin{pmatrix}
0 & m^{-1} \\ 
-m\Omega^2& -\gamma
\end{pmatrix}
\begin{pmatrix}
x_{0}\\
p_{0} 
\end{pmatrix}
+
\begin{pmatrix}
0\\ 
f
\end{pmatrix}
\end{align}
and for the perturbative terms:
\ba\label{eq:system2}
\begin{pmatrix}
\dot{\delta x}\\ 
\dot{\delta p}
\end{pmatrix}
=
\begin{pmatrix}
0 & m^{-1} \\ 
-m\Omega^2& -\gamma 
\end{pmatrix}
\begin{pmatrix}
\delta x\\
\delta p 
\end{pmatrix}
+
\begin{pmatrix}
4Ap_{0}^3/3m\\ 
-4A \gamma p_{0}^3/3
\end{pmatrix}
\end{align} 
In the following we will be integrating these equations of motion starting from a thermal state at $t=0$ and we will make the choice $\delta x(0)=0=\delta p(0)$.

We will assume the oscillator is driven only by noise arising from thermal and radiation pressure. The radiation pressure force obeys 
$\langle f_{\rm rad}(t')f_{\rm rad}(t'')\rangle = \hbar^2 G^2\alpha^2e^{\frac{-\kappa |t'-t''|}{2}}.$ See
\ref{steady} for details on the derivation. The thermal force obeys $\bra f_{T}\left(t'\right)f_{T}\left(t''\right)\ket=2k_B T\gamma m \delta\left(t'-t''\right)$, where $k_{B}$ is Boltzmann's constant.  The analysis can be simplified further if the optical cavity decay rate is sufficiently large $\kappa \gg  G\alpha, \Omega, \gamma$. Then the radiation contribution may also be approximated well by a white noise $\bra f_{rad}\left(t'\right)f_{rad}\left(t''\right)\ket=\left(4\hbar^2 G^2 \alpha^2/\kappa\right)\delta\left(t'-t''\right) $. The average number of photons in the cavity is $n_{\rm cav}=\alpha^2=4P/\kappa\hbar\omega_c$. We can also express the optomechanical coupling for a Fabry-Perot cavity as $G=\omega_c/L$ where $L$ is the length of the cavity. Thus we can re-express the radiation pressure white noise coefficient as $16\hbar\omega_c P/L^2\kappa^2$.

When the mechanical system is in steady state we wish to study the power spectrum of position fluctuations 
\begin{equation}
S_{xx}(\omega)= \int_{-\infty}^\infty \langle x(t)x(0)\rangle e^{i\omega t}dt. 
\end{equation}
Since we are studying a quantum mechanical theory, $ [ x(t), x(0)]\neq 0$ and consequently $S_{xx}(\omega)\neq S_{xx}(-\omega)$. We will mainly focus on the symmetrised noise spectrum $S(\omega)=[S_{xx}(\omega)+S_{xx}(-\omega)]/2$ \cite{clerk2010introduction,aspelmeyer2014cavity}. This is because the noise spectrum of an optomechanical position is just $S(\omega)$ plus a contribution due to the optical shot noise, see for example \cite{aspelmeyer2014cavity}.

As a result we need to calculate
\ba
\bra x\left(\tau\right) x\left(0\right)\ket&=\bra \left[x_{0}\left(\tau\right)+\delta x\left(\tau\right)\right]\left[x_{0}\left(0\right) +\delta x\left(0\right)\right]\ket\notag\\
&\simeq \bra x_{0}\left(\tau\right)x_{0}\left(0\right) \ket+\bra \delta x\left(\tau\right)x_{0}\left(0\right) \ket
\label{eq:correlation}\end{align}
The second expression arises since $\delta x(0)=0=\delta p(0)$.  The term $\langle \delta x\left(\tau\right)\delta x\left(0\right)\rangle$ is neglected since it is second order in $A$. We can use solutions of the above systems of equations \eqref{eq:system1} and \eqref{eq:system2} to compute the required correlation functions. Detailed calculations are in the Appendix. In both terms the correlators are averaged over the steady state, which does depend on the modified Hamiltonian. However since $\delta x\left(\tau\right)$ is itself first order in $A$ it is sufficient to compute $\bra \delta x\left(\tau\right)x_{0}\left(0\right) \ket$ over the unmodified steady state.

In the next sections we analyse the noise spectrum of the oscillator to find constraints on $\beta_{0}$.

\subsection{Noise spectrum}\label{Noise spectrum}

We find that the first-order modification to the oscillator noise spectrum may have significant contributions from both terms in equation (\ref{eq:correlation}). As shown in the Appendix the first term in \eqref{eq:correlation}, a correlation function calculated over a steady state with Hamiltonian \eqref{EffectiveHamiltonian}, is the sum of the correlation function with the standard commutation relation and a contribution that is proportional to $\hbar^2$ when $k_BT'\gg \hbar \Omega$. Here $T'$ is the effective temperature of fluctuations that drive the cavity with contributions from the true thermal noise and the radiation pressure noise (such an equilibrium temperature exists provided that the relation common to most optomechanical experiments $\kappa\gg\gamma$ holds; see \ref{steady}). This higher order dependence on $\hbar$ means that this correction is smaller by a factor proportional to $(p_{\rm ZPF}^2/mk_BT')^2\Omega/\gamma$ than the correction due to the second term in  (\ref{eq:correlation}). Here $p_{\rm ZPF}^2/mk_BT'$ is just the ratio of the variance of the momentum in the oscillator ground state to the momentum variance in the thermal state, thus it is small at high temperature. But due to the presence of the quality factor $\Omega/\gamma$ the overall factor may be significant. This fact is not unique to this specific commutator. In the Appendix we derive, in the presence of a potential  $V=B\sum_{m,n}a_{mn}a^{\dagger m}a^{n}$ (which may emerge from a modification to the commutator) to first order in $B$ the modification to the correlation function due to the modified steady state ensemble alone.

With this we find an expression for the modification of the symmetrised spectrum of mechanical fluctuations $S(\omega)$, (the sum of the first two terms of equation \ref{standardspec}). We are mainly interested in the case of high-$Q$ oscillators $\gamma\ll\Omega$ and in this case the expressions simplify considerably. We will be interested in two experimentally relevant regimes of frequency. Firstly $\omega\gg \Omega$ the so-called free-mass limit where frequencies of interest are far above the resonance frequency as is relevant for example in aLIGO. Secondly we consider frequencies close to resonance $\omega\sim\Omega$ which is where high sensitivity optomechanical experiments are usually probed and the rotating wave approximation is valid. For simplicity, for all the equations in this section we assume  $\kappa \gg  G\alpha, \Omega, \gamma$ so that radiation pressure noise can be approximated by a white noise. In the plots in later sections of the paper, which refer to specific experiments, we use the general spectrum \eqref{generalspectrum}.

In the free-mass limit the modification to the noise spectrum for the optomechanical system is (through \eqref{eq:fulllarge1}):
\begin{eqnarray}
\delta S\left(\omega\right)&\simeq& 
 \frac{2A\gamma\hbar^2}{\omega^2}\left[8\left(\frac{k_BT'}{\hbar\Omega}\right)^2\left(\frac{\Omega}{\omega}\right)^2+\frac{    1}{3 }\right]
\label{eq:fulllarge}\\
k_BT'&\simeq& \frac{8h\nu\mathcal{F}^2P}{\pi^2c^2\gamma m}+k_{B}T \label{eq:efftemp}
\end{eqnarray}
where   $\mathcal{F}=\Delta\omega_{FSR}/\kappa=\pi c/\kappa L$ is the finesse of the cavity. The dependence of $\delta S$ on the various parameters changes depending on which of the noise sources contributing to $k_BT'$ dominates.

The most important aspect here is that the spectrum has a different overall shape to the standard mechanical white noise spectrum in this frequency regime: $S\left(\omega\right)\propto \omega^{-4}$, for viscous/constant damping. The standard noise is inversely proportional to fourth power of driving frequency, as in the first term of the additional noise here (at lower frequencies outside the free-mass limit this takes a different shape to standard noise), but the second term from quantum zero-point energy is proportional to $\omega^{-2}$. This sort of effect could provide strong bounds on the minimal length. Even for the frequency-dependent `structural' damping of aLIGO (explored in the next section) which can be modelled by $\gamma=\Omega^2/\omega Q$ as explained in \ref{ligomodelling} the spectral frequency dependence is different to the standard expression.

We will now comment on how $\delta S(\omega)$ depends on the various parameters of the system. Specifically we will work in terms of the oscillator mass $m$, frequency $\Omega$, $Q$-factor, which for viscous damping is $Q=\Omega/\gamma$, temperature $T$, and optical power $P$. 

At frequencies $\omega/\Omega\gg k_{B}T'/\hbar\Omega$ the second term of the additional spectrum is dominant. In this regime it is inversely proportional to quality factor (keeping the resonance frequency constant). The damping model significantly affects the dependence on resonance frequency: for viscous damping it is proportional to $\Omega$ but for structural damping it is proportional to $\Omega^2$. In contrast to the behaviour of $S(\omega)$, $\delta S(\omega)$ is independent of mass. However, as we have noted earlier, in the version of the modified commutator theory in which $\beta_0$ is proportional to $\beta_{c}$ for the elementary components of the system and $A$ also depends on the number of constituent particles in the system, the noise expressed in terms of $\beta_{c}$ will depend on mass. We will explore this topic in more detail in the next section. None of the other parameters affect the noise appreciably.

In the other extreme, $\omega/\Omega\ll k_{B}T'/\hbar\Omega$, we may distinguish between dominant thermal or radiation pressure noise scenarios. If thermal noise dominates radiation pressure noise the additional spectrum is proportional to the square of the temperature, the dependence on all other parameters is the same as the case where the second term of additional spectrum is dominant.
On the other hand if radiation pressure noise dominates thermal noise then the additional noise increases quadratically with the power $P$ of the input laser and its frequency $\nu$, unlike the standard commutator radiation pressure noise which has a linear scaling with these parameters. But as in the standard expression it is inversely proportional to the square of mass in this frequency regime. The total noise is approximately proportional to the quality factor when radiation pressure noise is dominant. For viscous damping $\delta S(\omega)$ is proportional to $\Omega^{-1}$ but for structural damping it is proportional to $\Omega^{-2}$.

Explicitly the relative contribution of this noise to the standard thermal and radiation pressure noise is:
\ba\label{eq:relativelarge}
\frac{\delta S\left(\omega\right)}{S\left(\omega\right)}&\simeq
 \beta_{0}\frac{m}{(M_{P}c)^2}\left(8k_{B}T'+\frac{\hbar^2\omega^2}{3k_{B}T'}\right)
\end{align}
When the first term is dominant (again if $\omega/\Omega\gg k_{B}T'/\hbar\Omega$) this ratio is independent of frequency, except in the case of dominant radiation pressure noise and structural damping when it is proportional to $\omega$. Otherwise it rises as $\omega^2$, so to obtain a better bound on $\beta_{0}$ it is favourable to have a probe frequency away from the resonance frequency. However at $\omega\lesssim\omega_{SQL}$, where $\omega_{SQL}$ is the angular frequency where the standard quantum limit noise in ordinary quantum mechanics is reached, the measurement shot-noise (which we assume to be unaffected by the modified commutation relation studied here) dominates the mechanical noise. We account for this in the next section.

An assessment can be made of the relevance of the second term of equation \eqref{eq:relativelarge}. From the functional dependence of the equation it is clear that the optimal temperature that can practically be achieved to bound $\beta_0$ ,via a bound on the relative noise, for a system given $\omega$ is either $T_+$ (the maximum achievable temperature for the experiment) or  $T_-$ (the minimum achievable temperature consistent with the high temperature condition $k_{B}T'\gg\hbar\Omega$ which was used to derive the equation). Then if the first term at $T_+$ is much larger than the second term at $T_-$ the optimal bound will be at $T_+$ and only the first term is relevant for the experiment. This will be the case provided that $\omega/\Omega\ll\sqrt{(k_{B}T_+/\hbar\Omega) (k_{B}T_-/\hbar\Omega)}$, which is usually true in the high temperature limit.  

We now turn to consider the second, near resonance, regime where $\omega \simeq \Omega$. It is instructive to look specifically at $\omega=\Omega$ where the additional spectrum takes the form:
\ba
\delta S\left(\Omega\right)&= \frac{2 A \hbar ^2}{3 \gamma }
\label{eq:fullresonance}\\
\frac{\delta S\left(\Omega\right)}{S\left(\Omega\right)}&= \beta_{0}\frac{\hbar^2\Omega^2m}{3(M_{P}c)^2k_{B}T'}\label{eq:relativeresonance}
\end{align}
We see that the additional noise near resonance, up to  the order we have considered in perturbation theory, comes purely from zero-point energy of the oscillator. It is proportional to the quality factor and to $\Omega^{-1}$. The relative noise however is independent of the quality factor and favours a low temperature again.

It might be expected that the best place to probe $\delta S(\omega)$ is on the side of the resonance $\omega\simeq\Omega\pm\gamma''/2$ where $\gamma''=\Omega/Q$. Close to this frequency the absolute gradient of the standard power spectrum is highest so it is to be expected that the sensitivity to perturbation of spectrum is also maximum here. Indeed for high $Q$ the ratio between the first term of the perturbed spectrum \ref{eq:dSfull} and the sum of the standard thermal and radiation pressure spectra is maximum here.  We have:   
\ba
\delta S\left(\Omega\pm\frac{\gamma''}{2}\right)&\simeq \pm A(k_B T')^2 \frac{4}{\gamma^2 \Omega}
\label{eq:fullside}\\ 
\frac{\delta S\left(\Omega\pm\frac{\gamma''}{2}\right)}{S\left(\Omega\pm\frac{\gamma''}{2}\right)}&\simeq\pm  \beta_{0}\frac{4m\Omega k_BT'}{ \gamma \left(M_{P}c\right)^2}\label{eq:relativeside}
\end{align}
The main observation here is that $ \delta S\left(\Omega\pm\gamma''/2\right)\simeq6Q(k_{B}T'/\hbar\Omega)^2\delta S\left(\Omega\right)$. This relative noise always dominates the first term of the relative noise in the free-mass limit and also the second term if $\hbar\omega/k_{B}T'\ll  Q^{1/2} $.

In the regime where $\omega$ is comparable to $\kappa$ it is required to use the general formula \ref{standardspec} for the perturbed spectrum, provided that $\omega$ is not close to other resonances of the cavity. In subsequent sections where these formulae are referred this can be assumed.

\section{Constraints on the Modified Commutator}

In order to obtain constraints on the modified commutator for the oscillator's elementary components via experimental constraints on the noise spectra such as in previous section, parametrised in terms of $\beta_{0}$, we must rely on a model that relates the commutator for the components with that for its centre of mass.  Though an elementary component may have a simple commutation relation like \eqref{eq:beta} this may not be the case for the composite system. In \cite{quesne2010composite} a direct approach is taken, showing that for a many-component system where each component satisfies \eqref{eq:beta} with modification parameter $\beta_0$ it is possible to write a Hamiltonian for the centre of mass position and momentum where the modification parameter $\beta_0$ is rescaled. In general there are additional terms, not present without the modified commutator, that couple the total momentum to the relative momenta of the components.  Indeed, the terms involving relative momenta may dominate e.g. in ultra-hot bodies \cite{amelino2013challenge}.  This means generally the commutator structure is not universal; it requires information about microscopic degrees of freedom. A similar conclusion was drawn in \cite{pikovski2012probing} and \cite{amelino2013challenge}. 

To clarify the implication for the minimal length, a $\beta_0$ bound on the commutator obtained for the centre of mass of a system implies that the possible position variance of systems of that scale has a minimum value determined by $\beta_0$. An underlying assumption for associating a $\beta$ with a scale is that the minimum variance does not depend on the nature of the system apart from a concept of system size. If there is a minimum position variance associated to $\beta$ of systems of one scale it is smaller on larger systems that are a combination of systems of the original scale, though the exact relation also depends on the interrelations of the original system e.g. relative velocities. This is in contrast to standard quantum mechanics according to which the minimum position variance is exactly zero at all scales. 

In our scenario we are considering a rigid body at sufficiently low temperature and small internal motion that we neglect all relative momentum terms, which allows us to interpret that $\beta_{0}$ scales in proportion to $1/N^2$ if all N-components are identical. This implies that for an oscillator made of a single chemical element i.e. it is composed of $N_{a}$ atoms with mass $m_{a}$ then $\beta_{0}\sim \beta_{a}/N_{a}^2=\beta_{a}m_{a}^2/m^2$ where $\beta_{a}$ is the modification parameter for the individual atoms. But if the components are non-identical e.g. components are the nucleons and electrons of the oscillator then the relationship will be different. In the (uncharged) oscillator the number of electrons $N_{\rm elec}$ is less than the total number of nucleons $N_{\rm nuc}$, there are 3 quarks (of similar mass) per nucleon so $\beta_{\rm nuc}\sim \beta_{\rm quark}/3^{2}$, and assuming that the modification parameters of quarks and electrons are the same, being the most elementary particles currently known, $\beta_{\rm quark}=\beta_{\rm elec}=\beta_{e}$. From these assumptions and the fact that the mass of each nucleon $m_{\rm nuc}\simeq 1.67\times10^{-27}$kg$\simeq 2000 m_{\rm elec}$ it follows that, as shown in \cite{quesne2010composite}:
\ba\label{eq:be}
\beta_{0}&=N_{\rm nuc}\beta_{\rm nuc}\left(\frac{m_{\rm nuc}}{m}\right)^3+N_{\rm elec}\beta_{\rm elec}\left(\frac{m_{\rm elec}}{m}\right)^3\\
&\simeq\frac{\beta_{\rm nuc}}{N_{\rm nuc}^2}\\
&\simeq\frac{\beta_{e}m_{\rm nuc}^2}{9m^2}
\end{align}
Then a constraint on $\beta_{0}$ can be used to constrain $\beta_{e}$ and compared with previous constraints on it e.g. via measurements of Lamb shift in hydrogen atoms. 

Due to the model dependence of parametric relationship between different scales it is relevant in our analysis to optimise both bounds on $\beta_{0}$ and $\beta_{e}$ inferred from a given experiment. Consequently we will consider how bounds on both these parameters depend on the properties of the system, assuming that $\beta_0\propto \beta_e/m^2$.

\subsection{Estimating current constraints}

In this section we will consider how direct observations of optomechanical systems could place bounds on the modified commutator. We will consider two distinct regimes. Firstly the observation of noise in high precision interferometers like the Advanced Laser Interferometer Gravitational-Wave Observatory (aLIGO) in which the position fluctuations in the free mass limit are relevant. Secondly optomechanical experiments with nanomechanical systems where the position fluctuations of oscillators close to resonance can be observed that are due to both thermal and radiation pressure noise. These two example systems also operate in very different regimes for the mass of the oscillator. We will study the behaviour of $\delta S(\omega)$ as interpreted in the previous section using parameters drawn from recent experiments in these two regimes in order to get a quantitative idea of how big these bounds could be expected to be. There would be numerous additional technical issues in achieving these bounds in practice, not least because a realistic system would have more complex dynamics than the simple harmonic motion we consider in this paper. 

The aLIGO is a Michelson interferometer in which a high power laser beam is split into two 4km long arms of the interferometer and reflected back to interfere at the output port. Extremely small relative displacements of the arms, due to gravitational radiation for example, can be detected by shifts in the light intensity at the output but the sensitivity of this measurement is constrained by radiation pressure noise on the reflecting mirrors and shot noise at the output as well as the thermal and seismic noise on the mirrors and their suspending fibres \cite{martynov2016sensitivity}\cite{buonanno2001quantum}\cite{kimble2001conversion}.  

An accurate model of aLIGO is quite a lot more complicated than the one we have considered in this paper. Nevertheless it is of interest to understand whether there are potentially interesting bounds on $\beta_0$ arising from the measured aLIGO noise spectrum, by matching effective parameters for aLIGO to our simpler model. In the Appendix \ref{ligomodelling} we provide a translation between parameters of the multi-mirror apparatus of aLIGO to our single-oscillator single-cavity model; effective parameters of our model are displayed in table \ref{table:table2}. 

Some of the main issues in determining a set of parameters for our model that correspond to aLIGO are as follows. In aLIGO the mirrors of the optical cavities in each arm are designed such that $\kappa_{L}\sim\omega_{g}\gg \Omega$ where $\kappa_{L}$ is the decay rate of each cavity, $\omega_{g}$ is the order of magnitude of gravitational wave frequencies probed and $\Omega$ is the pendulum resonance frequency of the mirror-suspension system. Recall that in section \ref{Noise spectrum} we wrote simplified expressions for the perturbed spectrum in which the cavity dynamics are implicitly adiabatically eliminated. These expressions are valid only for frequencies $\omega \ll \kappa$ and when $\omega \sim \kappa$ we need to use the full expressions in the Appendix. In this regime, the radiation pressure noise spectrum is approximately inversely proportional to order sixth power of frequency\cite{martynov2016sensitivity}\cite{kimble2001conversion}\cite{chen2013macroscopic}, as in equation \ref{standardspec}. 
Moreover a `signal recycling' mirror in the output port of the aLIGO interferometer  produces correlations between radiation pressure and shot noise \cite{buonanno2001quantum}. Finally, leading contributions to thermal noise are internal to the suspension fibres and are usually associated with `structural' damping. As discussed in section \ref{Noise spectrum} a simple model for this is in terms of frequency-dependent damping of the form $\gamma=\Omega^2/ Q\omega $. This results in a spectrum inversely proportional to order fifth power of frequency unlike Brownian white noise \cite{saulson1990thermal}\cite{yamamotothesis}. The details of our model for structural damping are described in the Appendix. 
Thermal noise in the mirror coatings is also significant. Seismic noise is significant at frequencies below $\sim 10$ Hz, outside the frequency band of gravitational waves probed by aLIGO, and shot noise is dominant at frequencies above $\sim 100$ Hz. Finally, the noise spectrum for the more complicated optical system of aLIGO leads to a different balance of shot noise and radiation pressure noise contributions which we mimic by an effective detection inefficiency $\eta_2$.

The most recent observation of the total noise spectrum, shown in figure 7.3 of \cite{martynov2015lock} (also figure 5 in \cite{martynov2016sensitivity}), has a shape and order of magnitude consistent with a theoretical description that assumes standard canonical commutation relation and that these noise sources are independent, except between 20 and 100Hz there is a gap of order one magnitude between observed and expected noise which happens to be within the range of frequencies of gravitational waves probed as well as the segment where thermal noise dominates over shot noise. Via the conservative expression $\delta S/S_{obs}\left(\omega\right)\leq 1$, where $S_{obs}\left(\omega\right)$ is the observed spectrum, a tight constraint on $\beta_{0}$ is imposed which we can estimate by assuming that $\beta_0$ must be less than the value it would take for there to be some frequency at which $\delta S(\omega)=S_{obs}(\omega)$. 

Inserting the aLIGO parameters displayed in table \ref{table:table2} in the expression for $\delta S(\omega)$ \ref{generalspectrum} and comparing this to the measured value $S_{obs}(\omega)$ from figure 7.3 in \cite{martynov2015lock} results in the bound $\beta_{0}\lesssim 10^{21}$ if we are to satisfy $\delta S (\omega)\leq S_{obs}(\omega)$. Figure \ref{fig:LIGObounds} shows this rough bound as a function of probe frequency relative to resonance frequency:
\begin{figure}[h]
\centering
\includegraphics[scale=.6]{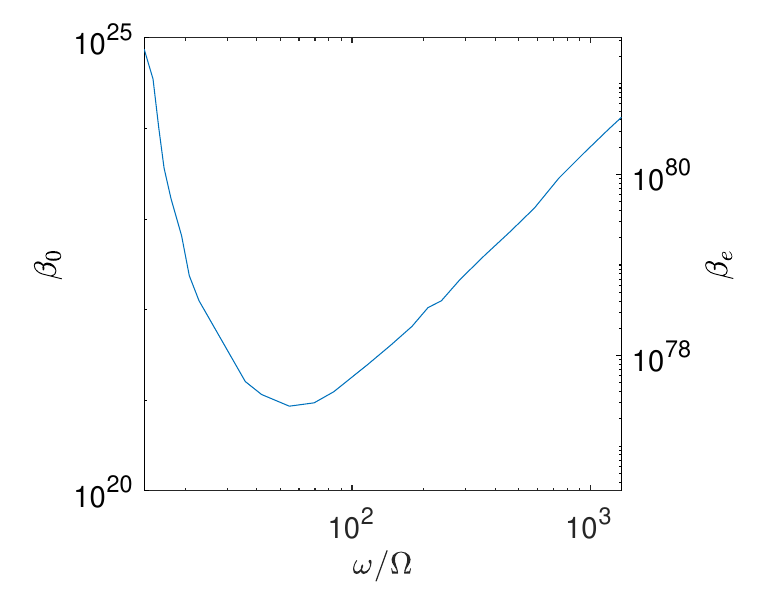}
\caption{Estimate of the bounds on $\beta$ arising from the measured noise spectrum in aLIGO experiments as a function of probe frequency. Here the upper bound on $\beta_{0}$ is derived at a given frequency by requiring that the predicted additional noise $\delta S(\omega)$ from equation \eqref{generalspectrum}  is less than the observed noise power spectrum at aLIGO, see main text for details. The $\beta_{e}$ parameter is the $\beta$ associated with elementary particles, inferred from $\beta_{0}$ according to equation \eqref{eq:be}.}
\label{fig:LIGObounds}
\end{figure}

As per table \ref{table:table1} this is comparable to the bound obtained in \cite{bawaj2015probing} through a tabletop experiment on macroscopic harmonic oscillators. It would appear both of these are amongst the tightest experimental bounds obtained so far but as discussed earlier it is only unambiguous to compare bounds that refer to the same types of system. The model discussed in the previous section associated with equation \eqref{eq:be} implies a bound for elementary particles $\beta_{e}\lesssim 10^{76}$ which is a much weaker constraint than that from Lamb shift measurements. 

\begin{table}[h!]
\begin{adjustwidth}{-.5in}{-.5in}
\caption{Effective experimental parameters and $\beta$ bounds}
\label{table:table2}
\begin{center}
\begin{tabular}{|l|l|l|l|l|l|l|l|}
\hline
\textbf{Experiment}                            & {aLIGO (2016) \cite{martynov2016sensitivity}\cite{martynov2015lock}} & { Purdy et.al (2013)\cite{purdy2013observation}} & { Teufel et. al (2016)\cite{teufel2016overwhelming}}  \\ \hline
{T (K)}                                 & 3.00E+02    & 1.70E-03               & 4.00E-02                                              \\
{m (kg)}                                & 1.00E+01 (reduced)    & 7.00E-12                & 8.50E-14                                                \\
{$\Omega$ (Hz)}                                & 4.15E+00    & 9.75E+06                & 5.88E+07                       \\
{$\gamma$ (Hz)}                                & 1.00E-06 (@ $\omega=\Omega$)    & 8.98E+03                & 1.53E+02                     \\
{Q}                                     & 1.33E+09    & 1.08E+03                & 3.83E+05                       \\
{$\nu$ (Hz)}                               & 2.82E+14    & 2.82E+14                & 6.71E+09                                               \\
{$L$ (m)}                                 & 4.00E+03    & 5.10E-03                & 4.00E-08                                   \\
{$\kappa$ (Hz)}                               & 4.78E+03 
   & 5.59E+06                & 6.64E+07                     \\
{$P$ (W)}                              & 3.60E+03 & 9.40E-05                & 7.80E-09                                        \\
{$\mathcal{F}$}                                     & 4.92E+01 
   & 3.30E+04                & 3.55E+08                                     \\
{Smallest S($\omega$) ($m^2/Hz$)} & 9.00E-40    & 4.00E-32                & 1.00E-26                                   \\
{$\beta_{0} [\beta_{e}]$ upper bound @ $\omega=\Omega$}                 & -   
 & 1E+41 [1E+73]               & 1E+42 [1E+70]                                                                   \\
{$\beta_{0}[\beta_{e}]$ upper bound @ $\omega=\Omega\pm\frac{\gamma''}{2}$}                  & -   
& 1E+31[1E+64]               & 1E+28[1E+57]                       \\
\hline               

\end{tabular}
\end{center}
\end{adjustwidth}    
\end{table}

Alternatively we can estimate the bounds that arise from optomechanical and electromechanical experiments with nanomechanical oscillators in which it has recently become possible to observe the effect of radiation pressure fluctuations. In the experiment of \cite{purdy2013observation} a silicon nitride membrane oscillator is placed in a Fabry-Perot cavity and driven at resonance by the radiation pressure fluctuations of a `signal' laser beam. A second much weaker `meter' laser beam at the same frequency is also passed through the cavity and its intensity photocurrent is measured. This photocurrent contains an imprint of the position fluctuations of the oscillator, and for sufficient measurement strength the fluctuations due to radiation pressure noise of the signal beam override those from the thermal Brownian motion of the oscillator. As a result the radiation pressure noise power spectrum can be measured cleanly. Another experiment \cite{teufel2016overwhelming} involves a microwave cavity optomechanical circuit containing a superconducting inductor resonating with a parallel-plate capacitor, i.e. a LC circuit in which the capacitor electrodes separation determines the resonance frequency of the microwave cavity. A coherent microwave drive at the cavity resonance is applied via a feed-line, the reflected beam is separated via a microwave circulator and the phase quadrature measured. The power spectrum of this signal is proportional to the mechanical spectrum of the capacitor, and again by optimising the measurement strength the radiation pressure fluctuations dominate the thermal noise. Both these experiments involve a single optomechanical cavity as in our setup so their parameters may be straightforwardly inserted in our equations. 

 To obtain the bounds near resonance we note first that in these particular experiments $\kappa$ is of the same order of magnitude as $\Omega$. Thus we are again required to use the exact formula \ref{generalspectrum} to estimate the perturbed spectrum. In both the experiments the observed position noise spectra are consistent with the standard noise from thermal, radiation pressure and shot noise \ref{standardspec}, assuming viscous damping, and so we may find rough estimates of bounds on $\beta$ in the way accomplished for aLIGO. Substituting the parameters in table \ref{table:table2} and comparing with the recorded spectrum in the experiments, the optimal bounds at resonance are $\beta_{0}\lesssim 10^{41}$, $\beta_{e}\lesssim10^{73}$ for experiment \cite{purdy2013observation} and $\beta_{0}\lesssim 10^{42}$, $\beta_{e}\lesssim 10^{70}$ for experiment \cite{teufel2016overwhelming}. Significant improvement of the bounds occur at $\omega=\Omega+\gamma''/2$, as described in the conclusion of section \ref{Noise spectrum}. For experiment \cite{purdy2013observation}  $\beta_{0}\lesssim 10^{31}$, $\beta_{e}\lesssim10^{64}$ is obtained and $\beta_{0}\lesssim 10^{28}$, $\beta_{e}\lesssim 10^{57}$ for the other experiment.
Even though tabletop experiment bounds of $\beta_{0}$ are far worse than aLIGO (which we have explored in the free-mass limit) the corresponding $\beta_{e}$ are better due to the use of low-mass oscillators. The equation \ref{eq:relativeside} suggests that observations on a system with parameters close to those of aLIGO but made for frequencies close to the mechanical resonance would improve bounds on $\beta_{0}$, due its higher mass, high mechanical quality-factor, and room-temperature setup.

\subsection{Future constraints}

We analyse how to optimise future experimental parameters to find better constraints on $\beta_{0}$ (and inferred value $\beta_{e}$) in the modified canonical commutation relation \ref{eq:beta}. In the previous section we derived constraints via bounds on the ratio between the observed noise on an oscillator and the additional mechanical noise on it due to the modified commutation relation (`relative noise criterion'). Here we will derive potential future constraints in a similar way except instead of substituting observed noise we use the theoretical expression for the noise level, which is just the sum \eqref{standardspec} of the standard mechanical and shot noise spectra. We also examine the experimental scenario where a fixed upper bound can be placed on the additional noise across frequencies. As in the previous section we choose to focus on two frequency regimes: $\omega\gg\Omega$ where the oscillator is treated as a free-mass, as well as the frequency just off its resonance frequency $\omega=\Omega+\gamma''/2$. 

Both large and small mass oscillators are of interest in future optomechanical experiments. In the previous section we examined aLIGO as an example of the former scenario but for the latter we explore here the experiment of \cite{murch2008observation} as a test-bed, which first reported an indirect signature of quantum radiation pressure noise.
In this experiment an ultracold rubidium atomic gas is confined by optical trapping in a Fabry-Perot cavity and continuously driven by a probe laser. The quantum radiation pressure noise of the laser heats the gas and via the observation of its evaporation rate the level of noise is inferred. This noise was found to be in agreement with standard quantum theory. If in a similar setup a direct observation of the mechanical position spectrum is possible, similar to the optomechanical experiments \cite{purdy2013observation}\cite{teufel2016overwhelming} that were considered in the previous section, our equations may be applied straightforwardly to obtain constraints on $\beta$ values. Some relevant parameters of this experiment are: $T=0.8\times10^{-6}$K, $m\simeq10^{-22}$kg, $\Omega=2\pi\times4.2\times10^{4}$Hz, $Q=42$, $P\simeq5.02\times10^{-13}$W, $\nu\simeq3.84\times10^{14}$Hz, $\kappa=2\pi\times6.6\times10^{5}$Hz and $L=1.94\times10^{-4}$m. Note also the experiment of \cite{brooks2012non}, not treated here, which also reported indirect observation of radiation pressure noise on an ultracold gas.

\begin{figure}[h!]
	\includegraphics[scale=0.45]{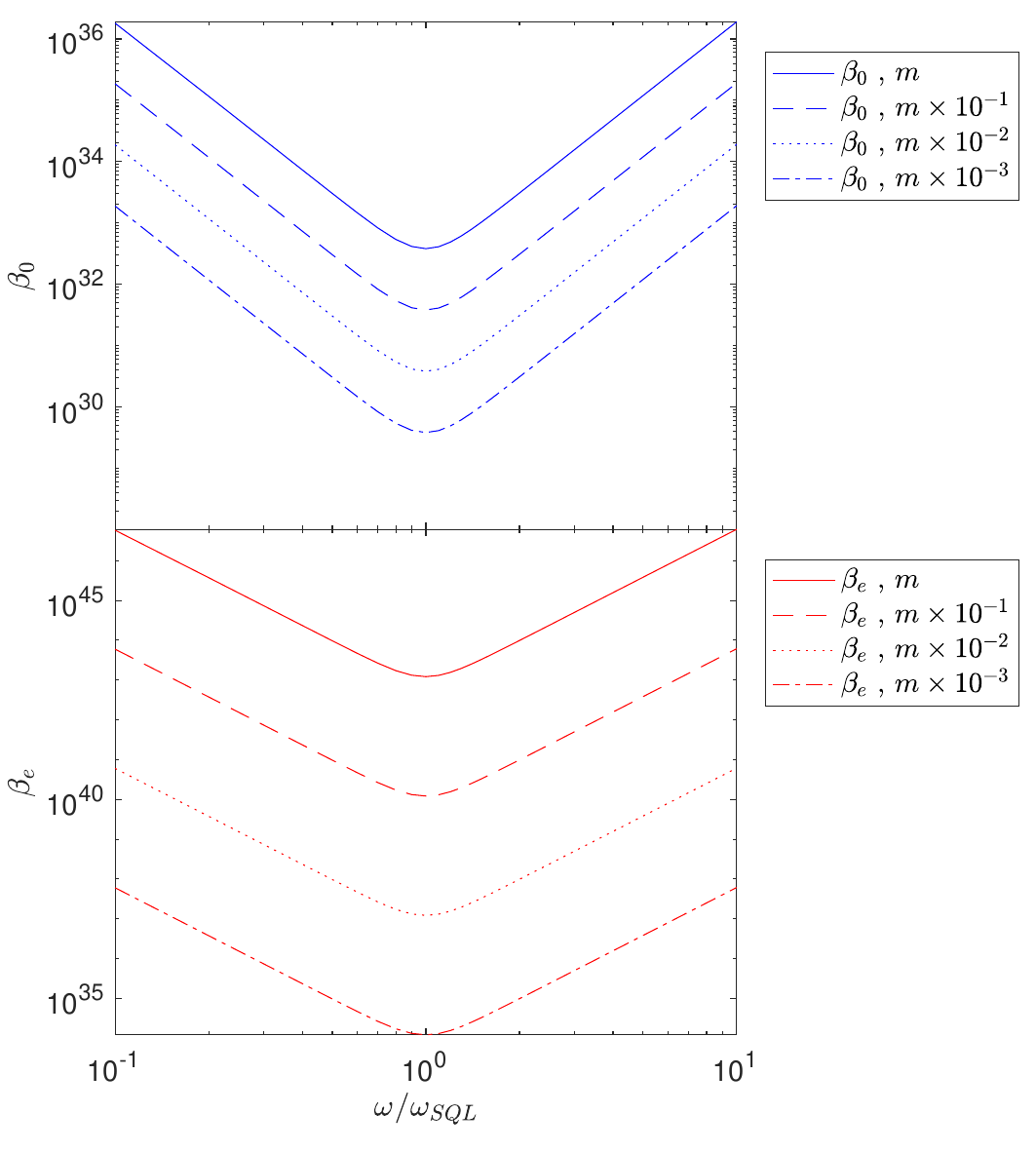}
	\includegraphics[scale=0.45]{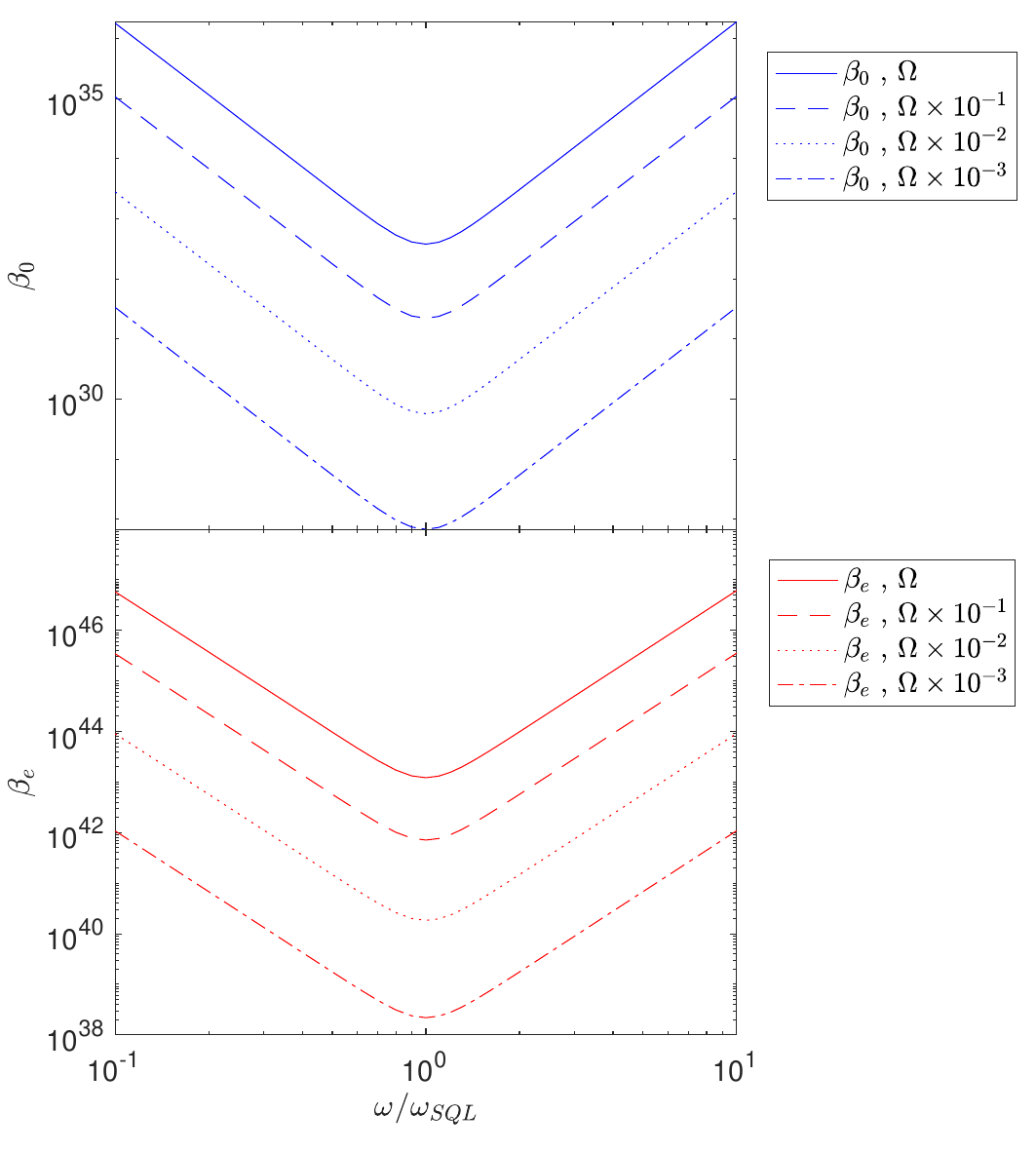}
	\includegraphics[scale=0.45]{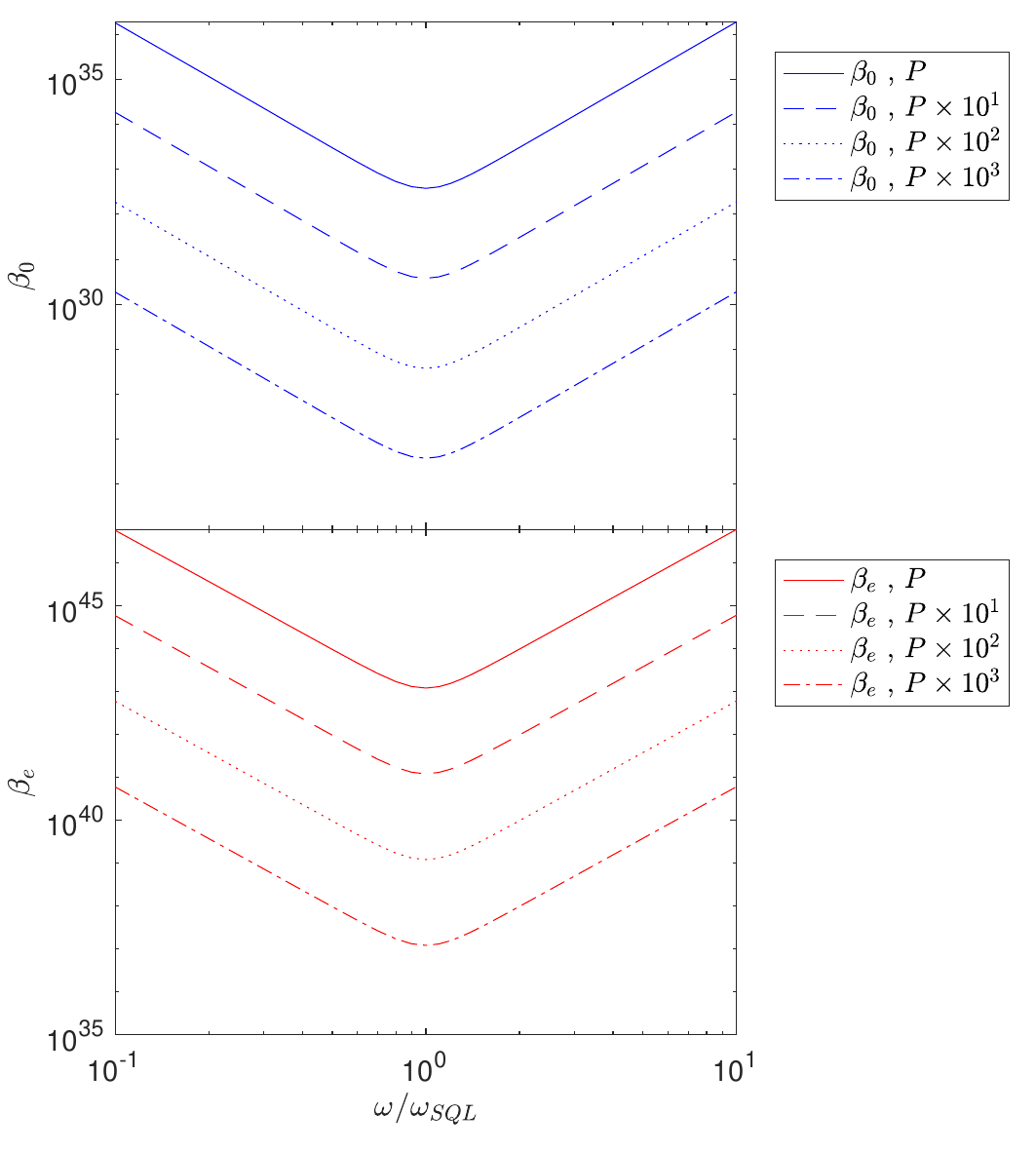}
	\caption{$\beta$ bounds vs. relative frequency ${\omega}/{\omega_{SQL}}$ for different masses, resonance frequencies and powers with respect to parameters taken from the experiment of Murch et. al \cite{murch2008observation} ($m\simeq10^{-22}$kg, $\Omega=2\pi\times4.2\times10^{4}$Hz and $P\simeq5.02\times10^{-13}$W are the standard values). The bounds are derived similarly to figure 1 by the requirement that $\delta S(\omega)/S_\text{std}(\omega)\lesssim1$ where $\delta S(\omega)$ is the perturbation to the mechanical noise spectrum due to the modified commutators and $S_\text{std}(\omega)$ is the analytical sum of theoretical thermal, radiation pressure and shot noise spectra. We use the full expression for $\delta S(\omega)$ equation \eqref{generalspectrum}. The sharp rise in bounds occurs when shot noise is dominant from $\omega\simeq \omega_{SQL}$ given by equation \eqref{omegaSQL}.}
	\label{fig:Murch vs parameters}
\end{figure}

\begin{figure}[h!]
	\includegraphics[scale=0.45]{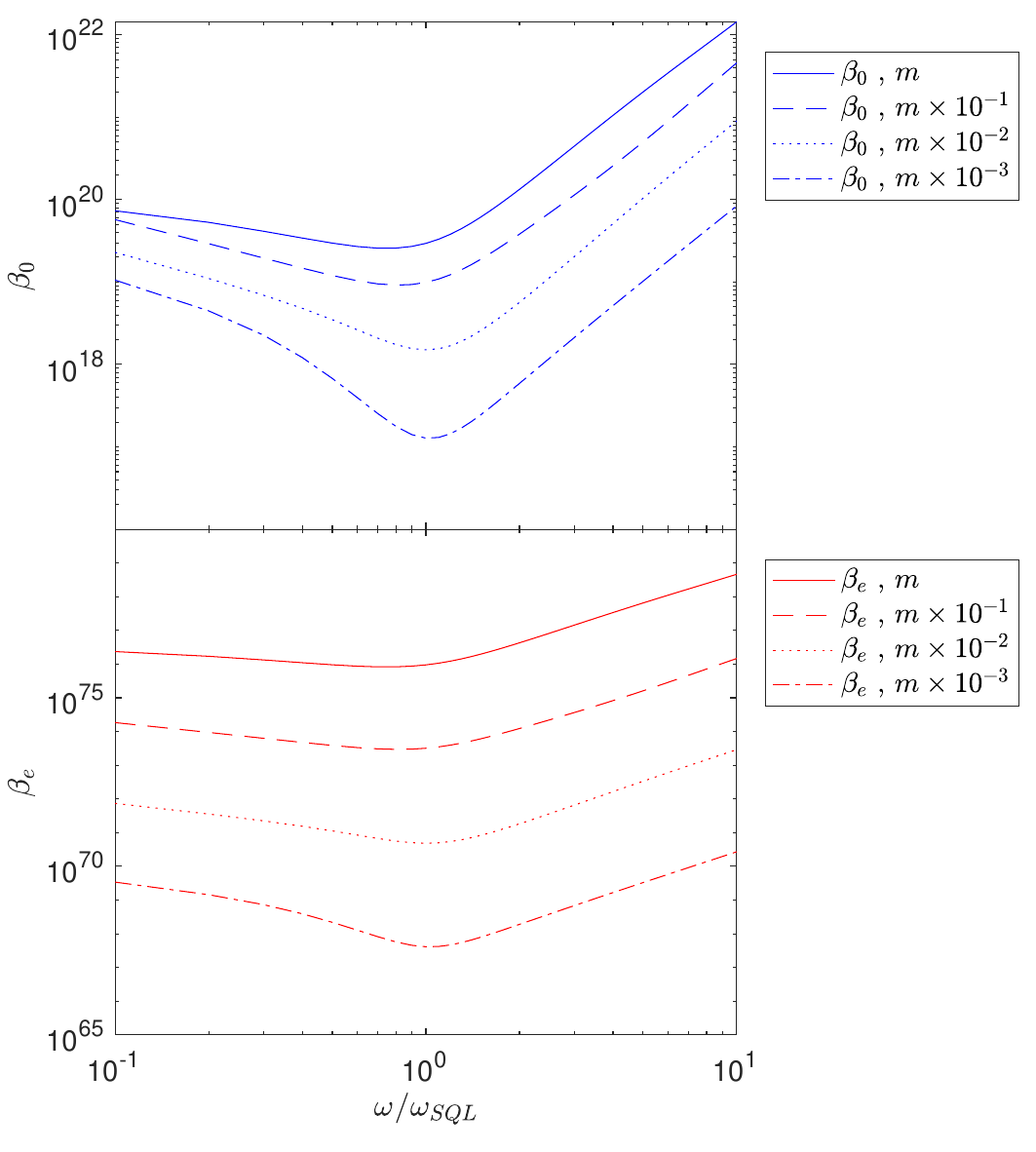}
	\includegraphics[scale=0.45]{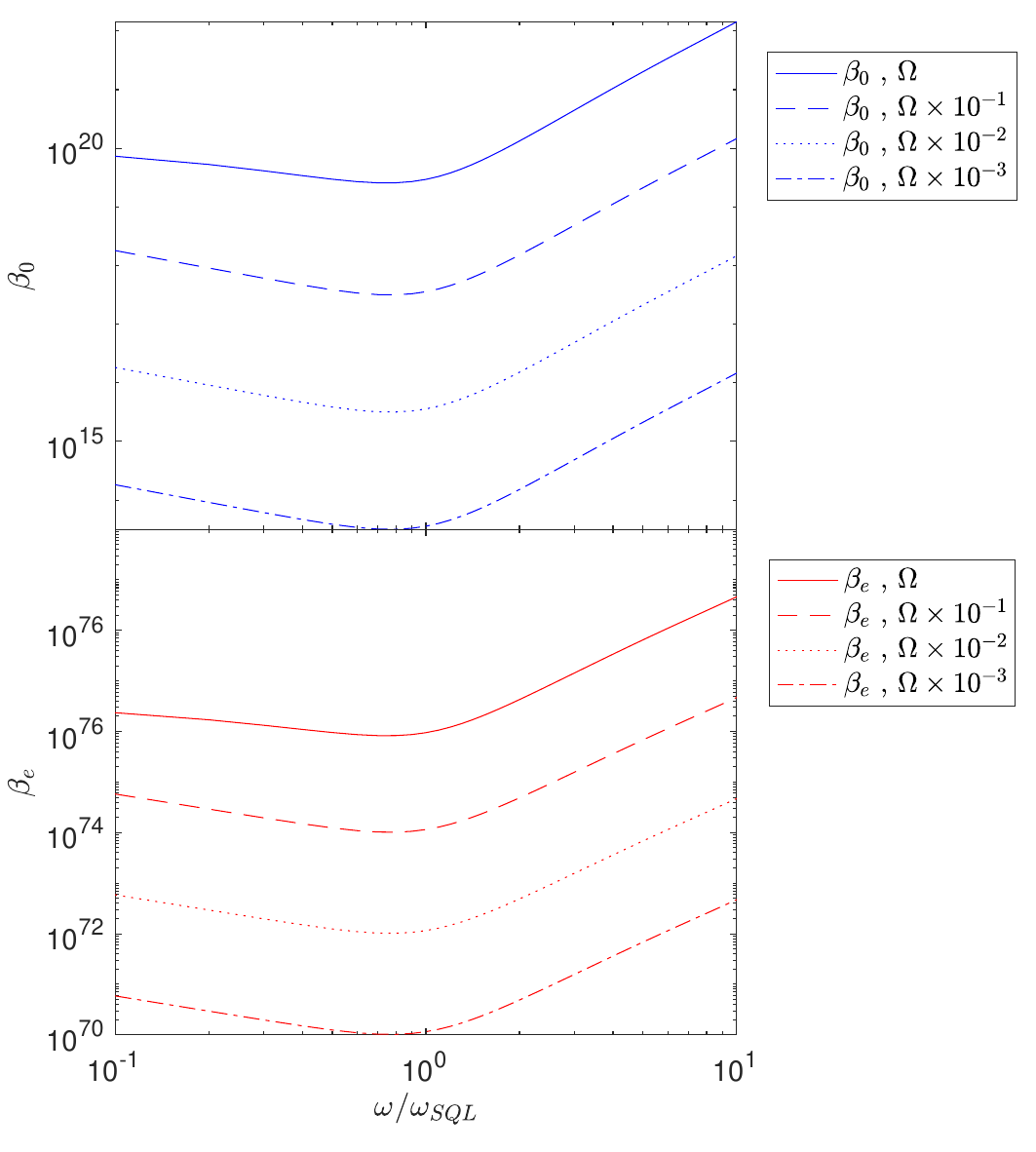}
	\includegraphics[scale=0.45]{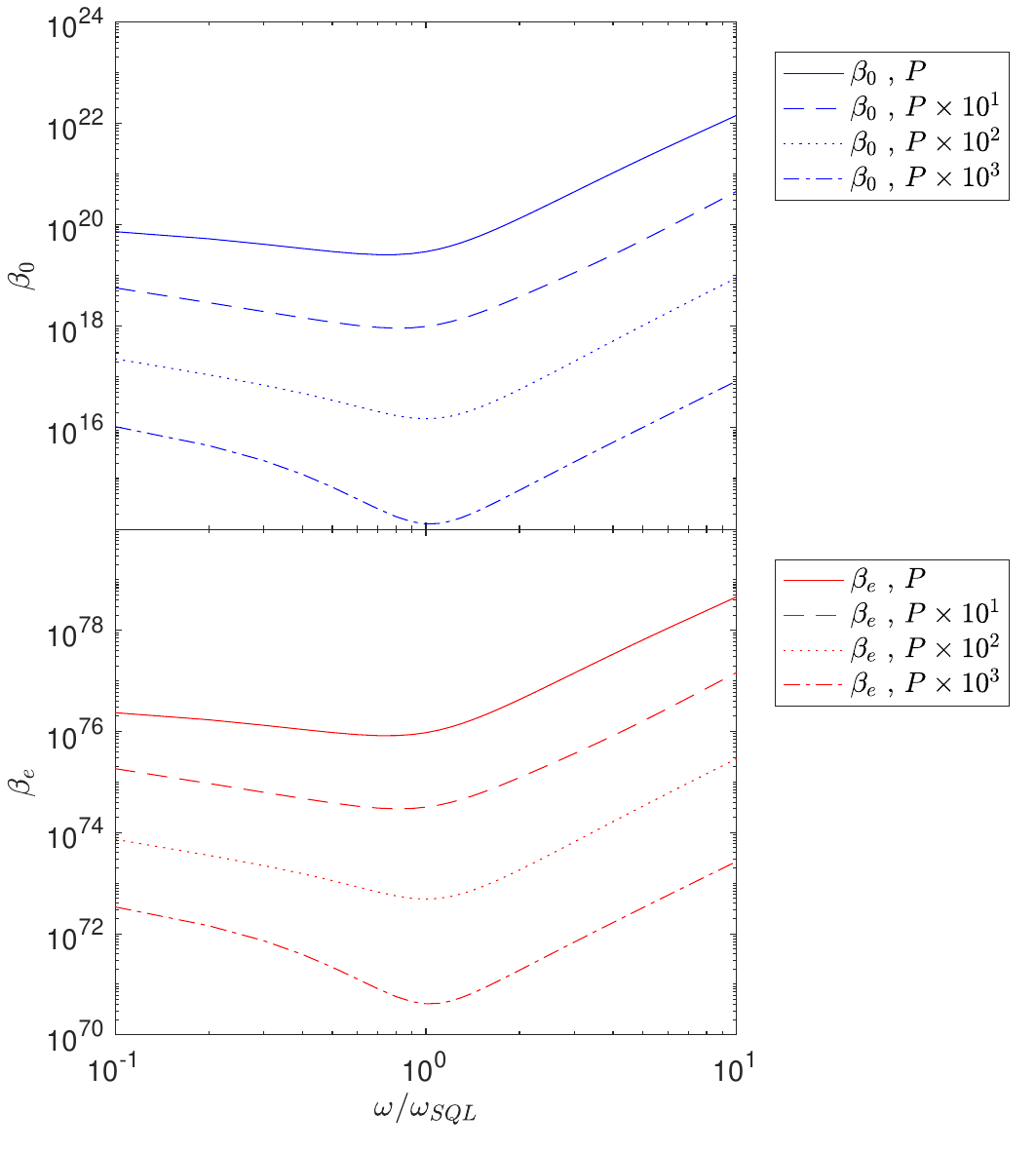}
	\caption{$\beta$ bounds vs. relative frequency ${\omega}/{\omega_{SQL}}$ for different masses, resonance frequencies and powers with respect to parameters taken from those of aLIGO \cite{martynov2016sensitivity} ($m\simeq10$kg, $\Omega\simeq2\pi\times6.6\times10^{-1}$Hz and $P\simeq3.7\times10^{2}$W are the standard values). The bounds are derived in the same way as those of figure 2. Once again the best bounds are obtained at $\omega\simeq \omega_{SQL}$. }
		\label{fig:LIGO vs parameters}
\end{figure}

The plots in figures \ref{fig:Murch vs parameters} and \ref{fig:LIGO vs parameters} show $\beta$-constraints attainable in parameter regimes similar to those of the experiment of \cite{murch2008observation} and aLIGO respectively when probed in the free-mass limit. Here we have chosen to adjust either $m, \Omega$ (keeping $Q$ constant) or $P$ individually in each of the plots by up to three orders of magnitudes from nominal values taken from those experiments. At each frequency shown the corresponding value of $\beta_0 (\beta_e)$ is found via the condition that $\delta S(\omega)/S_\text{std}(\omega)\lesssim1$. We must use the full expression \ref{generalspectrum} for $\delta S(\omega)$ as the adiabatic approximation is violated at higher frequencies, and $S_\text{std}(\omega)$ is taken to be the sum of thermal, radiation pressure noise and shot noise in equation \ref{standardspec}. We express angular frequencies relative to the standard quantum limit (SQL) frequency at which the minimum value of $S_\text{std}(\omega)$ is reached in the theory with $\beta=0$. The plots show that decreasing the mass or resonance frequency of the oscillator, or increasing the laser input power, improves the bounds with respect to $\omega/\omega_{SQL}$ at all frequencies considered.


 We stress that $\omega_{SQL}$ also varies with both these parameters:
\ba\label{omegaSQL}
\omega_{SQL}\simeq\sqrt{\frac{1}{8} \left(-\kappa ^2+\sqrt{ \left(\kappa ^2+4 \Omega ^2\right)^2+\frac{1024 \pi  \nu  P}{{L^2 m}}}+4 \Omega ^2\right)}
\end{align}
If we had plotted with respect to $\omega$ then a different parametric dependence holds which in the adiabatic regime obeys \ref{eq:relativelarge}  e.g. then higher (lower) mass is preferred for bounding $\beta_{0} (\beta_{e})$. Across our plots the bounds decrease till $\omega\simeq \omega_{SQL}$ after which they sharply increase. Thus the optimal bounds occur at this frequency. The initial decreasing behaviour in the case of plots associated with experiment of \cite{murch2008observation} can be attributed to the frequency dependence of the cavity susceptibility in the standard radiation pressure noise (see \ref{generalspectrum}), which makes an important contribution in the frequency regime we have considered, but in the plots associated with aLIGO this behaviour is due to the frequency dependence of structural damping. For aLIGO the plots corresponding to $m\times 10^{-3}$ and $P\times 10^{3}$ have a sharper drop to their minima, again due to the frequency dependence of cavity susceptibility in this parameter regime where $\kappa\sim\omega_{SQL}$. In both experiments the sharp increase in $\beta$ bounds is due to the domination of shot noise over mechanical noise from $\omega_{SQL}$. 

\begin{figure}[h!]
\centering
\includegraphics[scale=0.5]{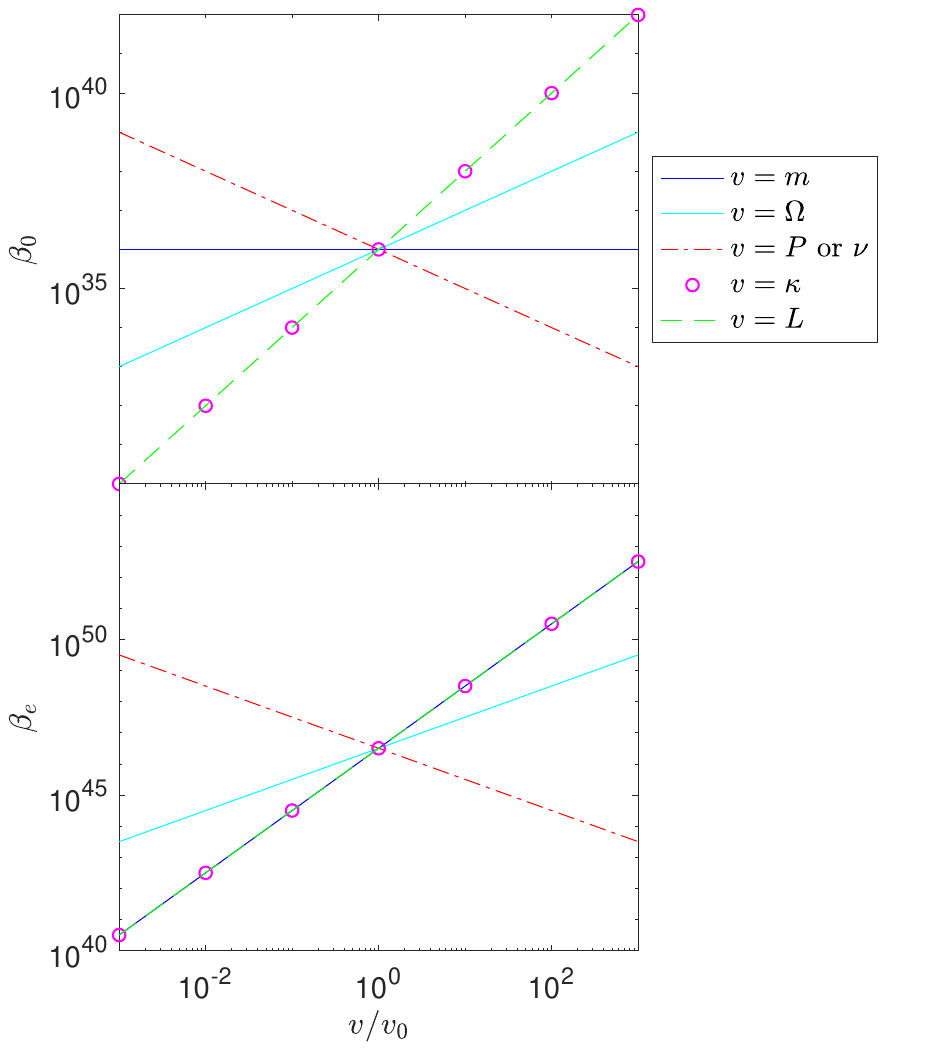}
\includegraphics[scale=0.5]{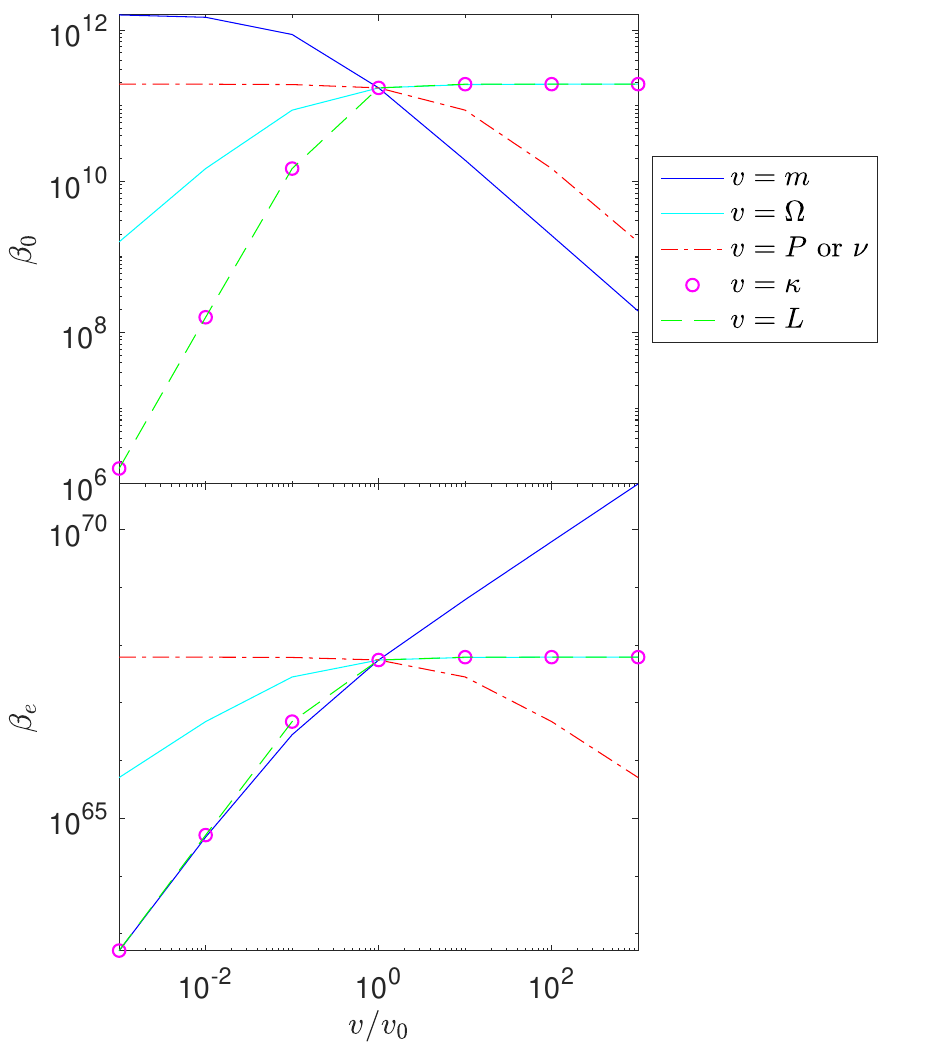}
\caption{Best bounds achievable at $\omega=\Omega+\gamma''/2$ by rescaling the value $v$  of a given variable, in the low-mass parameter regime based on Murch et. al \cite{murch2008observation} (left) or  the high-mass regime based on aLIGO \cite{martynov2016sensitivity} (right). The standard values $v_0$ are as in previous figures, except we set the central $\Omega$ value to $\Omega_0=\Omega_{i}\times10^{-3}$ where $\Omega_{i}$ is the original $\Omega$ value. $\nu\simeq3.85\times10^{14}$Hz, $\kappa=2\pi\times6.6\times10^{5}$Hz,$L=1.94\times10^{-4}$m in the low-mass parameter regime and $\nu\simeq2.82\times10^{14}$Hz,$\kappa\simeq5.55\times10^{2}$Hz,$L=4\times10^{3}$m in the high-mass regime. Here the bounds are derived from the relative noise as in previous figures.} 
\label{fig:FutureSideAll}
\end{figure}

\begin{figure}[h!]
\centering
\includegraphics[scale=0.5]{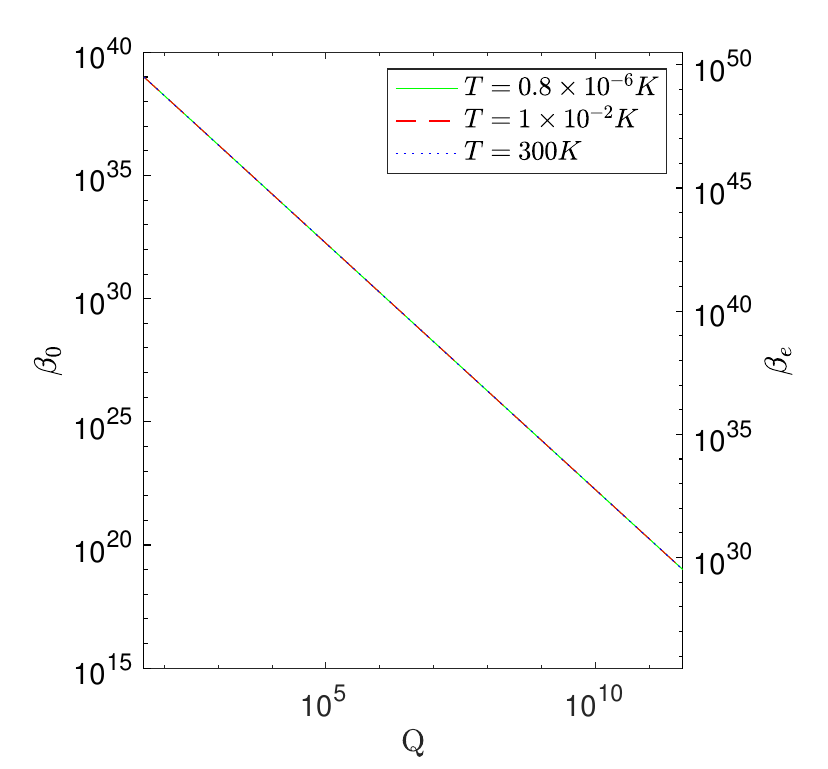}
\includegraphics[scale=0.5]{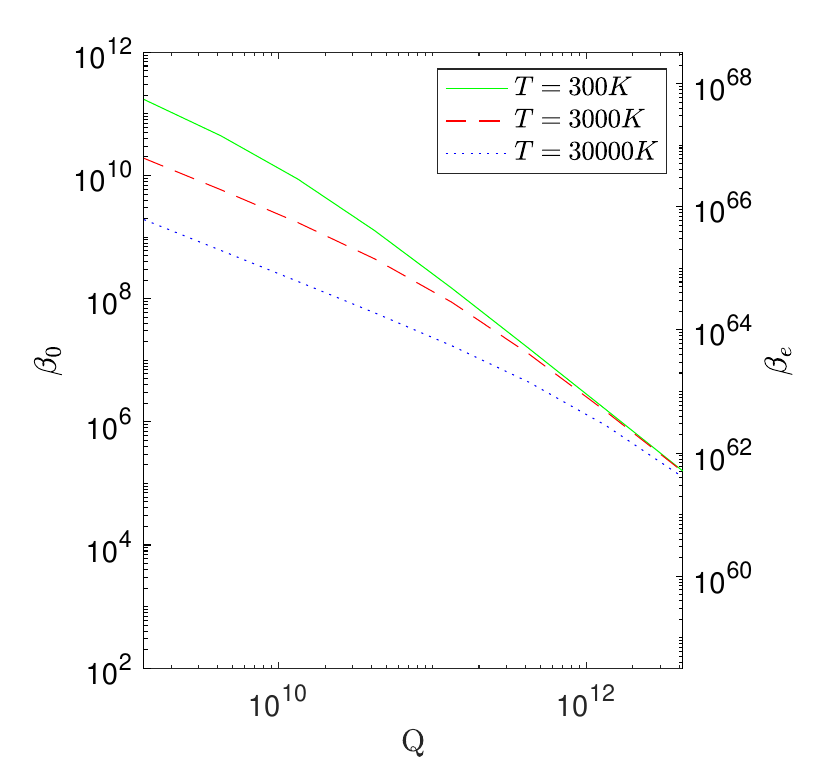}
\caption{$\beta$ bounds at $\omega=\Omega+\gamma''/2$ vs mechanical quality factor and varying thermal temperature, keeping other variables fixed at the values in the low-mass parameter regime based on Murch et. al \cite{murch2008observation} (left) and in the high-mass parameter regime based on aLIGO \cite{martynov2016sensitivity} (right). The bounds are derived similarly to those in the previous figure. The smallest quality factors considered here are the original values in the experiments. Increasing quality factor decreases the upper bounds in both parameter regimes but due to the high radiation pressure noise relative to thermal noise in the low mass parameter regime even an increase in over 8 orders of magnitude in temperature produces insignificant change in bounds in that case.}
\label{fig:Sidebounds}
\end{figure}

Looking close to resonance at $\omega=\Omega+\gamma''/2$ we obtain for both parameter regimes the potential $\beta$ bounds displayed in figures \ref{fig:FutureSideAll} and \ref{fig:Sidebounds}. In the first figure we vary by three orders of magnitude in both directions oscillator mass $m$, laser power $P$ (which has same effect as varying $\nu$), cavity decay rate $\kappa$ and length $L$ around the values in the original experiments. Each parameter is varied independently, so $\kappa$ is varied with $L$ fixed (and vice versa) which implies the cavity finesse is also varied, and varying $L$ keeping $\nu$ fixed implies varying the optical mode number. Also the oscillator resonance frequency $\Omega$ is varied up to six orders of magnitude below the original experimental values (this is shown in the plots as varying by three orders of magnitude in both directions around $\Omega_{0}=\Omega_{i}\times10^{-3}$ where $\Omega_{i}$ is the original experimental value). While varying $\Omega$ we keeping the mechanical $Q$ constant. We choose this range of values for $\Omega$ in order that $\kappa\gg\gamma$ which is required for the approximations we use in deriving these equations to be valid, see the Appendix. In the other figure we show the dependence on the oscillator $Q$ (keeping $\Omega=\Omega_{i}$ constant) and thermal temperature $T$. Both figures show a contrast in the dependence on variables between the two experiments. 

We see that for an experiment like that in \cite{murch2008observation} the greatest improvement in bounds are from individually increasing the quality factor or reducing the cavity length or decay rate (if probing $\beta_{e}$ then reducing the mass has the same effect); for every order of magnitude adjustment of these variables both $\beta_{0}$ and $\beta_{e}$ are reduced by approximately two orders of magnitude. The apparent overlap of the plots varying cavity length and decay rate comes from the fact that over all the specific parameter space considered here the system remains in the adiabatic regime. Here the dependency of the $\beta$ bounds on $\kappa$ and $L$ can be completely expressed in terms of cavity finesse $\mathcal{F}\propto 1/\kappa L$.  This is generally not the case. Increasing the laser power or frequency or reducing oscillator resonance frequency are also preferred though here the bounds change by only one order of magnitude for every order of magnitude adjustment of these variables. The high sensitivity to optical parameters is a consequence of dominant radiation pressure noise in this experiment, but as a result altering the mass or thermal temperature has negligible effect on the $\beta_0$ bound. In the plots associated with aLIGO better bounds are obtained by adjusting the variables in the same way but in contrast the bounds do not vary appreciably for lower power and higher cavity length and decay rate, and better $\beta_{0}$ bounds arise from increasing the mass and thermal temperature. This difference in behaviour is from the fact that in that parameter regime corresponding to aLIGO radiation pressure noise does not dominate thermal noise.

 We examine how to reach specific new bounds on the modified commutation relation. Firstly, comparison of the figures suggests that generally for low mass oscillators similar to \cite{murch2008observation} probing in the free-mass limit is favourable but for experiments in parameter regimes more similar to those of aLIGO it is better to observe close to resonance. To reach $\beta_e\lesssim 10^{37}$, just one order of magnitude off the bound inferred from Lamb shift \cite{das2008universality}, one could for example probe an experiment like \cite{murch2008observation} in the free-mass limit whilst increasing the laser power by approximately three orders of magnitude or reducing the mass (resonance frequency) by two (four) orders of magnitude from the nominal values of the experiment.  Note that tuning a combination of these parameters would require less adjustment of the magnitudes. This bound could also be reached when probing close to resonance by, for example, lowering any two of the oscillator mass, cavity length or decay rate by three orders of magnitude (increasing the power by same amount has a weaker effect) or increasing the oscillator quality factor quite significantly to $Q\simeq10^7$. For an experiment like aLIGO we see from the plots that $\beta_0\lesssim10^{14}$ could be reached in the free-mass limit by lowering $\Omega$ by three orders of magnitude. But probing close to resonance it is possible, extrapolating figure \ref{fig:FutureSideAll} , to reach the target value $\beta_0\lesssim1$ by also similarly lowering cavity length and decay rate. 
 
The above bounds on $\beta$ values are derived via a constraint on the relative noise, but if an experiment places a fixed upper constraint on the additional noise itself then optimal parameters have a different scaling and tighter bounds could be achieved. In the adiabatic regime the equation \eqref{eq:fulllarge} is the relevant additional noise in the free-mass limit. As discussed earlier, if thermal noise is dominant then larger $T$ is preferred as well as smaller $Q$ (though to derive the additional spectrum $Q\gg 1$ is assumed). But once radiation pressure noise is dominant then to significantly improve bounds on both $\beta$ values larger $Q$ and optical parameters $\mathcal{F}, P, \nu$ or smaller $m$ and $\Omega$ are required. As an example, in the experiment of \cite{murch2008observation} if the additional noise across frequencies is bounded by the noise at the SQL frequency so $\delta S(\omega)\lesssim S_{std}\left(\omega_{SQL}\right)\sim 10^{-29}$m$^{2}$/Hz, then $\beta_{e}\lesssim 10^{38}$ can be obtained at $\omega=10\Omega$, a few orders of magnitude away from the bound from Lamb shift.
For an experiment like aLIGO we can similarly deduce $\beta_{e}\lesssim 10^{74}$. By lowering its reduced mass significantly to $m=10^{-9}$kg and keeping the noise constrained to be below the noise at the SQL frequency (corresponding to that mass), we obtain a $\beta_{e}$ bound comparable to that from Lamb Shift and also $\beta_{0}\lesssim1$. Bringing the mass down further to $m=10^{-20}$kg allows to potentially reach $\beta_{e}\lesssim 1$. Probing at $\omega=\Omega+\gamma''/2$ equation \eqref{eq:fullside} applies in the adiabatic regime and implies large Q is preferred particularly when radiation pressure noise is dominant. When the additional noise can be constrained by the SQL noise then for aLIGO we have already have the strong bound $\beta_0\lesssim10^{-14}$ and by lowering (increasing) the mass (quality factor) by two orders of magnitude we can bound $\beta_e$ better than from Lamb shift. The bound $\beta_{e}=1$ could be reached by adjusting the mass(quality factor) by another ten (nine) orders of magnitude. For the experiment of \cite{murch2008observation} with a similar noise constraint in fact we have $\beta_{e}\lesssim10^{31}$, and $\beta_{0}=1$$  (\beta_{e}=1)$ would be achieved by raising quality factor to $Q=10^6 (Q=10^9)$. 


\subsection{General driven oscillator}

The method we have focused on to place tight bounds on a modified canonical commutator involves constraints on the corresponding modified thermal Brownian noise and quantum radiation pressure noise on an oscillator, the key properties being that the modified noise relative to the standard noise is amplified with greater effective temperature and quality factor, and that the modified noise spectrum has a different scaling with respect to probe frequency. If $\beta_{0}$ ($\beta_{e}$) is probed a larger (smaller) mass is preferred when thermal noise is dominant, but either way a high effective temperature bath is also preferred.

 We can heuristically translate this dependency on effective temperature to a general out-of-equilibrium scenario by considering the oscillator driven by a stochastic approximately white noise force $f_{drive}$ associated with an effective temperature $T_{eff}$ that fulfils the equipartition relation: $k_{B}T_{eff}=m\bra v^{2} \ket$. Then the free-mass limit equation \eqref{eq:relativelarge} becomes:
$\beta_{0}\lesssim\left(M_{P}c\right)^2\left(8m^2\bra v^2\ket+\hbar^2\omega^2/2\bra v^2\ket\right)^{-1}$.
 If the mirrors in an experiment like aLIGO are moved so that $\bra v^2\ket\sim 5.3\times10^{-2}m^2s^{-2}$ then $\beta_{0}\lesssim1$ can be reached (neglecting the effect of probe frequency as we have assumed high effective temperature and taking the probe frequency to be sufficiently smaller than the frequency, following the discussion after equation \eqref{eq:relativelarge}. But if a higher probe frequency can achieved it could be advantageous). At $\Omega+\gamma''/2$, from equation \ref{eq:fullside} we have $\beta_{0}\lesssim (M_{P}c)^{2}/4Q m^{2}\bra v^{2}\ket$ so with the possibility of using high-$Q$ material like sapphire only $\bra v^2 \ket \sim 10^{-10} m^2s^{-2}$ may be required to reach $\beta_0\lesssim 1$. The bound on $\beta_{e}$ is independent of mass for a given speed. With a high-$Q$ oscillator and $
 \bra v^2 
 \ket=10^{-4} c^2$ the bound is approximately $\beta_{e}\lesssim10^{31}$, a tighter constraint on minimal length than inferred from Lamb shift and experiments at the Large Hadron Collider. For driven systems approaching higher speeds we expect even better bounds can be attained though this would enter the relativistic regime, which is outside the scope of our analysis, and a non-perturbative treatment should be used to gain a more accurate picture.

\section{Discussion}

We have shown that probing the mechanical noise on harmonic oscillators with high precision, as in recent optomechanical experiments, is a viable technique to constrain small modifications of quantum mechanics found in models of quantum gravity, specifically the canonical commutation relations. 
Based on a conservative criterion that constrains the perturbation of the standard mechanical spectrum by the observed spectrum, we inferred that currently a bound on $\beta_{0}$ via aLIGO is comparable with that from other experiments. We found that the strongest constraints on both $\beta_0$ and $\beta_e$ appear to arise from probing the spectrum close to the mechanical resonance frequency or close to where the standard quantum limit of noise is reached. Strikingly, by adjusting both optical and mechanical parameters by a few orders of magnitude it is possible to probe if $\beta_0\lesssim1$, which could be interpreted as a glimpse of the Planck scale in the GUP framework. New bounds on $\beta_e$ using our proposal may be achievable in the near term, which would be impressive since it would mean a better understanding of the mechanics of elementary particles like electrons can be achieved by measurements of macroscopic systems. As noted in the previous section, for a general noise source the ideal scenario is to drive the oscillator so it reaches large variation of speeds. High quality factor oscillators are also generally desirable, though this may not be the case if the $\beta$ bounds are derived from a fixed upper constraint on the additional noise.

Other proposals like that in \cite{pikovski2012probing} and experiments in \cite{marin2014investigation} \cite{bawaj2015probing}, and more recently \cite{bushev2019testing}\cite{bonaldi2020probing} have also studied macroscopic oscillators to probe modifications of commutation relations. It may seem odd that \cite{pikovski2012probing} provides a formula that generally favours small mass to constrain $\beta_{0}$ unlike our result and the other experiments. However this is not a fundamental contradiction as the proposal there utilises multiple optical pulses in order to probe an induced geometric phase, whilst here we are concerned with the modification of the position noise spectrum/auto-correlation function. Also, in contrast to our work, in other papers the behaviours of the mean or variance of position or energy of systems were studied. For example in \cite{bawaj2015probing} and more recently \cite{bushev2019testing} the commutator we have studied was mapped via the perturbed Hamiltonian to the classical solution for the mean position of a Duffing oscillator. The nonlinearity implies a consequent amplitude-frequency effect. This is also visible in the correlation function we have derived that incorporates, unlike previous papers, fluctuations from a noisy bath and both viscous and structural damping. We have also kept within a fully quantum formalism that shows potentially significant zero point energy contributions. The figure of merit for the amplitude-frequency effect turns out to be similar to that derived from equation \eqref{eq:fullside} for the frequency close to resonance, after applying the equipartition theorem. But in the previous papers the uncertainty in resonance frequency was significantly enhanced by the resolution bandwidth of the spectrum analyser and the intrinsic oscillator non-linearity (consequently only small amplitudes of motion were probed).  Here we expect over the whole frequency range of the spectrum that only close to resonance, as seen by application of the rotating wave approximation, may the intrinsic non-linearity obscure effects from modified commutators. However the experiments we have considered do not significantly show such limitations, so that the corresponding frequency uncertainty is taken to be the oscillator damping rate (hence the preference for a high quality factor).

 While this paper was in preparation \cite{bosso2018potential} also analysed modifications of radiation pressure noise (and shot noise) in aLIGO but instead chose to only modify the optical commutation relations. But there it is reported that aLIGO measurements already constrain the relevant modification parameters to far below 1. We focused on a modified commutator solely for the mechanical degrees of freedom of the system oscillator; this restriction can be justified as the standard quantum position and momentum operators usually apply to massive degrees of freedom (in the non-relativistic limit). Indeed we have shown how significant improvements could be found from mechanical effects alone especially in future tabletop experiments. A general treatment incorporating modified optical relations is outside the scope of our paper but it can easily be seen that the net modified mechanical position noise spectrum would be a linear combination of our results and the effects from solely modifying the optical relations, to first order in $\beta_0$ and any similar modification parameters for the optical fields. One key property of a purely mechanical treatment is that the shot noise is unmodified but in \cite{bosso2018potential} it is reported that the modified optical relations imply a term reducing this noise unless the optical field is squeezed to unrealistic amounts. 
 
  Also while this paper was in preparation there have been multiple reports of observation of quantum radiation pressure noise on room temperature oscillators \cite{sudhir2017quantum}\cite{purdy2017quantum}\cite{cripe2019measurement}. These new techniques are promising for future applications of our results, in particular the experiment of \cite{cripe2019measurement} consists of observations across broadband and off-resonance frequencies relevant to our analysis of the free-mass limit.

Here we mainly used a conservative bound on relative noise as the basis for our results, but in future studies other measures could give significant improvements. The relative noise approach is equivalent to assuming that the error on the noise spectrum is proportional to the total observed noise (where the factor is up to an order of magnitude). But as we have noted it could be that the precision measurements are associated with a fixed error that is independent of the observed noise, at least over a certain frequency band. Precise knowledge of the experimental error at individual frequencies would lead the way for tighter $\beta$ bounds. This could account for statistical features like the number of data samples taken in the experiment. A related point is that use of tailored measurement methods and/or squeezed light fields \cite{buonanno2001quantum}\cite{kimble2001conversion}\cite{caves1981quantum} in new interferometers, enhanced even more by quantum entanglement\cite{ma2017proposal}, can increase position measurement sensitivity beyond the SQL limit. This is particularly important for our work which shows that in the free-mass limit the optimal bounds are inferred at approximately the ordinary SQL frequency when accounting for shot noise. This paves the way for application of fundamental limits in parameter estimation like the quantum Cramer-Rao bound which has recently been studied for gravitational wave detectors \cite{tsang2011fundamental}\cite{miao2017towards}. Note that the modified commutation relation implies a shift of both the frequency and the noise level at the noise spectrum minimum.

Our study can also be extended to other interpretations of the minimal length. As mentioned earlier there are several other proposed modified commutation relations associated with quantum gravity, like equations \ref{mumodified} and \ref{deltamodified}. It turns out that in previous experiments tighter bounds are generally obtained for the parameters in these equations, so it can be expected that following our methods the same will be achieved and spectral features amplified where there is higher non-linearity. The commutator we have studied is however well-motivated from the point of view of maintaining a UV momentum cut-off \cite{kempf1995hilbert}. It can be associated with several variance-based uncertainty principles and for some measures of `minimal length' can be expressed via entropic uncertainty relations  as recently shown \cite{abdelkhalek2016optimal}. 

Here we have performed a perturbative treatment in $\beta_0$. This breaks down for $\beta_0 (p/M_P c)^2\gg1$, so related effects for high-energy oscillators remain to be examined.

In an upcoming publication we explore a general quantum-mechanical treatment of continuous position measurement in the presence of different modifications of the canonical commutation relation, revealing new features of measurement back-action potentially applicable to a wide spectrum of experimental scenarios including the setting in this paper \cite{girdhardohertypreparation}.

\textit{Acknowledgements} We acknowledge support from the  Australian Research Council (ARC)  via the
Centre of Excellence in Engineered Quantum Systems
(EQuS), Project No. CE170100009. P.G. thanks the organisers of the 10th Relativistic Quantum Information Workshop (2016) where parts of this research were first presented. A.C.D acknowledges very helpful conversations with Michael Vanner and Warwick Bowen on modified commutators and radiation pressure noise.


\appendix
\section{}
In this appendix we derive the modified spectrum of mechanical fluctuations from the solutions of the equations $\eqref{eq:system1}$ and $\eqref{eq:system2}$.

Equation (\ref{eq:system1}) describes a damped driven oscillator with frequency $\Omega$ and damping rate $\gamma$. We will consider only the underdamped regime $\gamma<2\Omega$ in which case eigenvalues of the linear system are  $\lambda_{\pm}=-\gamma_{0}\pm i\omega_{0}$ with:
\ba\label{eq:eigenvalues}
\gamma_{0}=\frac{\gamma}{2},\quad  \omega_{0}=\sqrt{4\Omega^2-\gamma^2}/2.
\end{align}

The solutions are
\ba\label{eq:solutions1}
\begin{pmatrix}
x_{0}\left(t\right)\\ 
p_{0}\left(t\right)
\end{pmatrix}=\frac{1}{2\omega_0}
\begin{bmatrix}
ie^{\lambda_{+}t}\left[x_{0}\left(0\right)\lambda_{-}-p_{0}\left(0\right)/m\right]+ia/m+\ {\rm h.c.}  \\ 
i\lambda_{+}  e^{\lambda_{+}t}\left[x_{0}\left(0\right)\lambda_{-}m-p_{0}\left(0\right)\right]-i\lambda_+a^*+ \ {\rm h.c.} 
\end{bmatrix}
\end{align}
and
\ba\label{eq:solutions2}
\begin{pmatrix}
\delta {x}\left(t\right)\\ 
\delta {p}\left(t\right)
\end{pmatrix}
=\frac{1}{2\omega_0}
\begin{bmatrix}
ie^{\lambda_{+}t}\left[\delta x\left(0\right)\lambda_{-}-\delta p\left(0\right)/m\right]+ic/m+ \ {\rm h.c.} \\ 
i\lambda_{+} e^{\lambda_{+}t}\left[\delta x\left(0\right)\lambda_{-}m-\delta p\left(0\right)\right]-i\lambda_+ c^*+ {\rm h.c.} 
\end{bmatrix}
\end{align}
where:
\ba
a=\int_{0}^{t} f\left(s\right)e^{\lambda_{-}\left(t-s\right)} ds,\quad c=\frac{4}{3}A \lambda_{-}\int_{0}^{t}p_{0}^3e^{\lambda_{-}\left(t-s\right)} ds
\end{align}

The force term $f$ is a sum of thermal and radiation pressure noise. The thermal noise is, in the high temperature limit, treated as a white noise obeying the correlation function $\bra f\left(s'\right)f\left(s''\right)\ket=2k_B T\gamma m \delta\left(s'-s''\right)$.  For radiation pressure noise the correlation function is:
\begin{equation}\label{eq:radcorrel}
\langle f_{\rm rad}(s')f_{\rm rad}(s'')\rangle = \hbar^2 G^2\alpha^2e^{-\kappa |s'-s''|/2}.
\end{equation}
See \ref{steady} for details on the derivation. The white noise approximation of this corresponds to adiabatic elimination of the cavity field, given by
 $\bra f_{\rm rad}(s')f_{\rm rad}(s'')\ket=(4\hbar^2 G^2 \alpha^2/\kappa)\delta\left(s'-s''\right)$.

\subsection{First term}

Assume the oscillator at initial time is in thermal equilibrium associated with quantum state:
\ba
\rho&=Z^{-1}e^{-\beta H}\notag\\
&=Z^{-1}e^{-\beta\left(H_{0}+V\right)}
\end{align}
where $H_{0}=m\Omega^2 x^2/2+\tilde{p}^2/2m$, $V=A\tilde{p}^4/3m$ and $\beta=1/k_BT'$. Now $\rho$ satisfies the Bloch equation:
\ba
\frac{\partial}{\partial \beta}\rho=-H\rho 
\end{align}
with $\rho\left(\beta=0\right)=I$.
We can use time-dependent Hamiltonian perturbation theory  to compute the modified thermal state and its correlation functions since this equation is equivalent, via a Wick rotation, to the evolution of a unitary operator in the Schr\"odinger picture for a system evolving under a time-dependent Hamiltonian. 
We keep only the first perturbation as we only keep terms first order in $A$. For any observable $O$ we then have:
\ba
\bra O \ket_{H}&=\tr\left(Z_{H}^{-1}e^{-\beta H}O\right)\notag\\
&\simeq Z_{H_{0}}{Z_{H}}^{-1}\left[\tr\left(Z_{H_{0}}^{-1}e^{-\beta H_{0}} O\right)-\tr\left(Z_{H_{0}}^{-1}e^{-\beta H_{0}} \left\{\int_{0}^{\beta}e^{\beta' H_{0}}Ve^{-\beta' H_{0}} d\beta'\right\}O\right)\right]
\end{align}
We use the subscript $H$ to emphasize that the expectation value is taken using the thermal state of the full Hamiltonian.

Substituting the identity operator for O above we get:
\ba
Z_{H_{0}}{Z_{H}}^{-1}&=\left(1-\bra\int_{0}^{\beta}e^{\beta' H_{0}}Ve^{-\beta' H_{0}} d\beta'\ket_{H_{0}}\right)^{-1}\notag\\
&\simeq 1+\bra\int_{0}^{\beta}e^{\beta' H_{0}}Ve^{-\beta' H_{0}} d\beta'\ket_{H_{0}}
\end{align}
Here we have introduced the notation $\bra O \ket_{H_{0}}=\tr\left(Z_{H_0}^{-1}e^{-\beta H_0}O\right)$ for expectation values according to the thermal state of the unperturbed Hamiltonian $H_0$.
Therefore to first order in A:
\ba
\bra O \ket_{H}=\bra O \ket_{H_{0}}+\bra O \ket_{H_{0}}\bra\int_{0}^{\beta}e^{\beta' H_{0}}V\left(\beta'\right)e^{-\beta' H_{0}} d\beta'\ket_{H_{0}}-\bra \left\{ \int_{0}^{\beta}e^{\beta' H_{0}}Ve^{-\beta' H_{0}} d\beta'\right\}O\ket_{H_{0}}
\end{align}

With these perturbation theory expressions in hand we are able to compute the first term in the correlation function in \eqref{eq:correlation}:
\ba\label{eq:correlationfirst1}
\bra x_{0}\left(\tau\right)x_{0}\left(0\right) \ket_{H}=&\bra x_{0}\left(\tau\right)x_{0}\left(0\right) \ket_{H_{0}}+\bra x_{0}\left(\tau\right)x_{0}\left(0\right) \ket_{H_{0}}\bra\int_{0}^{\beta}e^{\beta' H_{0}}Ve^{-\beta' H_{0}} d\beta'\ket_{H_{0}}\notag\\
&-\bra  \int_{0}^{\beta}e^{\beta' H_{0}}Ve^{-\beta' H_{0}} x_{0}\left(\tau\right)x_{0}\left(0\right)d\beta'\ket_{H_{0}}.
\end{align}

At this point we can use the solutions for $x_0(\tau)$ and $p_0(\tau)$. From now on we use the notation $x_{0}$ and $p_{0}$ for $x_{0}\left(0\right)$ and $p_{0}\left(0\right)$ respectively. We will also only compute the real part of $\bra x_{0}\left(\tau\right)x_{0}\left(0\right) \ket_{H}$ since this is all that will be required for the symmetrised spectrum which is our main interest and use the subscript $R$ to indicate the real part of a correlation. 
For a standard thermal state $\bra p_{0}x_{0}\ket_{H_{0},R}=0=\bra a x_{0}\ket_{H_{0}}$. Then:
\ba\label{eq:correlationfirst3}
\bra x_{0}\left(\tau\right)x_{0} \ket_{H_{0},R}=\frac{1}{2\omega_0}\bra x_{0}^2\ket_{H_0}\left(i\lambda_{-}e^{\lambda_{+}\tau}-i\lambda_{+}e^{\lambda_{-}\tau}\right).
\end{align}

In our scenario $H_{0}=\hbar\Omega\left(b^{\dagger}b+1/2\right)$ so $e^{\beta' H_{0}}pe^{-\beta' H_{0}}=-ip_{\rm ZPF}\left(b e^{-\beta'\hbar\Omega}-b^{\dagger}e^{\beta'\hbar\Omega}\right)$ where $p_{\rm ZPF}=\sqrt{\hbar\Omega m/2}$ is the zero-point momentum. 
Since we are averaging over a thermal state, which is a Gaussian state, we can apply Wick's theorem \cite{gardiner2004quantum}. For thermal states the only non-zero terms in $\bra\int_{0}^{\beta}e^{\beta' H_{0}}Ve^{-\beta' H_{0}} d\beta'\ket_{H_{0}}$ are those with equal powers of $b$ and $b^{\dagger}$ and we have $\bra b^{\dagger n}b^n\ket
=n!\bra b^{\dagger}b\ket^n$ \cite{thouless1957use}.
This implies:
\begin{equation}
\bra\int_{0}^{\beta}e^{\beta' H_{0}}Ve^{-\beta' H_{0}} d\beta'\ket_{H_{0}}= \left(\frac{Ap_{\rm ZPF}^4}{3m}\right)\left[3\beta\left(2\bra  b^{\dagger 2}b^2\ket+4\bra b^{\dagger}b\ket+1\right)\right]
=\frac{Ap_{\rm ZPF}^4}{mk_BT'x_{\rm ZPF}^4}\bra x_{0}^2\ket_{H_0}^2
\end{equation}
where $x_{\rm ZPF}=\sqrt{\hbar/2m\Omega}$ is the zero-point position.

We can proceed in the same way to show
\ba
\bra\int_{0}^{\beta}e^{\beta' H_{0}}Ve^{-\beta' H_{0}} x_{0}^2 d\beta'\ket_{H_{0}}
=&\bra\int_{0}^{\beta}e^{\beta' H_{0}}Ve^{-\beta' H_{0}}\left\{x_{\rm ZPF}^2\left(b^2+b^{\dagger 2}+2b^{\dagger}b+1\right)\right\} d\beta'\ket_{H_{0}}\notag\\
=&\frac{Ap_{\rm ZPF}^4}{m}\bra x_{0}^2\ket_{H_0}\left\{\frac{1}{2\hbar\Omega}\left[\left(\frac{\bra x_{0}^2\ket_{H_0}}{x_{\rm ZPF}^2}+1\right)^2\left(e^{-2\beta\hbar\Omega}-1\right)-\left(\frac{\bra x_{0}^2\ket_{H_0}}{x_{\rm ZPF}^2}-1\right)^2\left(e^{2\beta\hbar\Omega}-1\right)\right]\right. \notag \\
&\left.+\beta\left(3\left(\frac{\bra x_{0}^2\ket_{H_0}}{x_{\rm ZPF}^2}\right)^2-2\right)\right\}
\end{align}
We will mainly be interested in the high temperature limit $\hbar\Omega/k_BT'\ll 1$ in which case we can simplify this expression as follows
\begin{equation}
\bra\int_{0}^{\beta}e^{\beta' H_{0}}Ve^{-\beta' H_{0}} x_{0}^2 d\beta'\ket_{H_{0}}
\simeq \frac{Ap_{\rm ZPF}^4}{mk_BT'}\bra x_{0}^2\ket_{H_0}\left(\frac{\bra x_{0}^2\ket_{H_0}^2}{x_{\rm ZPF}^4}-\frac{4}{3}\right).
\end{equation}
Finally we can also show
\ba\label{eq:xpterm}
{\rm Re }\left(\bra\int_{0}^{\beta}e^{\beta' H_{0}}V\left(\beta'\right)e^{-\beta' H_{0}} p_{0}x_{0} d\beta'\ket_{H_{0}}\right)=0
\end{align}

So, substituting in 
\eqref{eq:correlationfirst1} we get
\ba\label{eq:pertcorrelation}
\bra x_{0}\left(\tau\right) x_{0}\ket_{H, R}\simeq \bra x_{0}\left(\tau\right)x_{0} \ket_{H_{0}, R}+\frac{2Ap_{\rm ZPF}^4}{3\omega_0 m k_BT'}\bra x_{0}^2\ket_{H_0} \left(i\lambda_{-}e^{\lambda_{+}\tau}-i\lambda_{+}e^{\lambda_{-}\tau}\right),
\end{align}
when $\tau\geq 0$. In the following sections we will compute these correlation functions for $\tau\geq 0$ only and infer the case of negative $\tau$ from general properties of the correlation functions.
This expression implies that the perturbation to the standard correlation function in equation \eqref{eq:pertcorrelation} is to leading order quadratic in $\hbar$, arising from the zero-point energy of the oscillator.

Let us briefly consider how this calculation will proceed for other modified commutation relations. For other commutation relations for which $[x,p]$ depends on $p$  it will be possible to define a momentum-like observable $\tilde{p}$ satisfying $[x,\tilde{p}]=i\hbar$ to first order. Writing the Hamiltonian $H$ in terms of $\tilde{p}$ we would find that it is perturbed by a ``potential'' $V$. A similar approach would work for those commutation relations for which $[x,p]$ depends on $x$. Any perturbation $V$ can be expressed in the form:
\ba
V=B\sum_{m,n}a_{mn}a^{\dagger m}a^{n}
\end{align}
where B is a small parameter. 
The interaction picture representation of $V$, $V_{I}$, is:
\ba
V_{I}=B\sum_{m,n}a_{mn}e^{i\left(m-n\right)\Omega t}a^{\dagger m}a^{n}
\end{align}
Performing a Wick rotation $it\rightarrow\hbar\beta$ we can find these expressions for arbitrary operator $O$:
\ba
\bra O \ket_{H_{0}}\bra\int_{0}^{\beta}V_{I}\left(\beta'\right) d\beta'\ket_{H_{0}}&=B\left(\frac{1}{\hbar\Omega}\sum_{m,n:m\neq n}a_{mn}\left(m-n\right)^{-1}\left(e^{\hbar\Omega\beta\left(m-n\right)}-1\right)\bra a^{\dagger m}a^{n}\ket\right.\notag\\
&+\left.\beta\sum_{n}a_{nn}\bra a^{\dagger n}a^{n}\ket\right)\bra O\ket_{H_{0}}\\
&=B\beta\sum_{n}a_{nn}\bra a^{\dagger n}a^{n}\ket\bra O\ket_{H_{0},th}
\end{align}
where the last line is true for a thermal state, and
\ba
\bra\int_{0}^{\beta}V_{I}\left(\beta'\right) d\beta'(O)\ket_{H_{0}}&=B\left(\frac{1}{\hbar\Omega}\sum_{m,n:m\neq n}a_{mn}\left(m-n\right)^{-1}\left(e^{\hbar\Omega\beta\left(m-n\right)}-1\right)\bra a^{\dagger m}a^{n}O\ket_{H_{0}}\right.\notag\\
&+\left.\beta\sum_{n}a_{nn}\bra a^{\dagger n}a^{n}O\ket_{H_{0}}\right)
\end{align}
From equations $\eqref{eq:correlationfirst3}$ and $\eqref{eq:xpterm}$, which is valid for general $V$ averaged over a thermal state, the contribution to equation \ref{eq:correlationfirst1} from the perturbed Hamiltonian is:
\begin{eqnarray}
\delta \bra x_{0}\left(\tau\right)x_{0} \ket_{H, R}&=&\frac{1}{2\omega_0}\left(i\lambda_{-}e^{\lambda_{+}\tau}-i\lambda_{+}e^{\lambda_{-}\tau}\right)\left(\bra x_0^2 \ket_{H_{0},th}\bra\int_{0}^{\beta}V_{I}\left(\beta'\right) d\beta'\ket_{H_{0},th}-\bra\int_{0}^{\beta}V_{I}\left(\beta'\right)x_0^2 d\beta'\ket_{H_{0},th}\right)\notag\\
&=&\frac{B}{ 2\omega_0 }\left(i\lambda_{-}e^{\lambda_{+}\tau}-i\lambda_{+}e^{\lambda_{-}\tau}\right)\left(\frac{1}{k_B T'}\sum_{n}a_{nn}\left(\bra a^{\dagger n}a^{n}\ket\bra x_0^2\ket_{H_{0},th}-\bra a^{\dagger n}a^{n}x_0^2\ket_{H_{0},th}\right)\right.\notag\\
&&\left.-\frac{1}{\hbar\Omega}\sum_{m,n:m\neq n}a_{mn}\left(m-n\right)^{-1}\left(e^{\hbar\Omega\beta\left(m-n\right)}-1\right)\bra a^{\dagger m}a^{n}x_0^2\ket_{H_{0},th}\right)
\end{eqnarray} 

Thus, for a general potential that may be associated with a modification to quantum mechanics, it is not necessary that this part of the modification to the correlation function vanishes in the high temperature limit.


\subsection{Second term}
In the previous section we calculated, at $t=0$, the perturbation to the thermal state due to the modified commutation relation. Working in the Heisenberg picture we take at this time that $\delta x\left(0\right)=\delta p\left(0\right)=0$ and now calculate the influence of the modified evolution of the position operators over time. Since all of these corrections will be proportional to $A$ we can evaluate expectation values using the thermal state of the unperturbed harmonic oscillator, so now $\bra x_{0}^2\ket$ will mean $\bra x_{0}^2\ket_{H_0}$. The second term of \eqref{eq:correlation}, from \eqref{eq:solutions2}, is:
\ba 
\bra \delta x\left(\tau\right)x_{0}\ket_{R}
&=\frac{1}{2m\omega_0}\bra \left[ic(\tau)-ic^*(\tau)\right]x_{0}\ket_{R}\notag\\
&=-\frac{2A}{3m\omega_0}\int_{t}^{t+\tau}\bra p_{0}\left(s\right)^3x_{0} \ket_{R}\left(i\lambda_{+}e^{\lambda_{+}\left(\tau-s\right)}-i\lambda_{-}e^{\lambda_{-}\left(\tau-s\right)}\right) ds.
\end{align}
And
\ba
p_0\left(s\right)
&=x_0j\left(s\right)+p_0g\left(s\right)+h\left(s\right),
\end{align}
where 
\ba
j\left(s\right)&=\frac{\lambda_{+}\lambda_{-}m}{2\omega_0} \left(ie^{\lambda_{+}s}-ie^{\lambda_{-}s}\right)\\
g\left(s\right)&=\frac{1}{2\omega_0}\left(i\lambda_{-}e^{\lambda_{-}s}-i\lambda_{+}e^{\lambda_{+}s}\right)\\
h\left(s\right)&=\frac{1}{2\omega_0}\left[i\lambda_{-}a\left(s\right)-i\lambda_{+}a^*\left(s\right)\right].
\end{align}

Now since the unperturbed oscillator is in a Gaussian thermal state we can use Wick's theorem to obtain symmetrically ordered moments (denoted by subscript $sym$) to get, for example $\bra x_0^2p_0^2\ket_{sym}=\bra x_0^2\ket\bra p_0^2\ket+2\bra x_0p_0\ket_{sym}^2$. Consequently,
\ba
\bra p_0^3\left(s\right)x_0\ket_{R}&=j\left(s\right)^3\bra x_0^4\ket+3g\left(s\right)^2j\left(s\right)\bra x_0^2p_0^2\ket_{sym}+3\bra h\left(s\right)^2\ket j\left(s\right)\bra x_0^2\ket\\
&=3\bra x_{0}^2\ket j(s)\left[j(s)^2\bra x_{0}^2\ket +g(s)^2\bra p_{0}^2\ket +\bra h(s)^2\ket\right]
\end{align}

So then:
\begin{eqnarray} 
\bra \delta x\left(\tau\right)x_{0} \ket_{R}
&=&-\frac{2A}{m\omega_0}\bra x_{0}^2\ket \left[\bra x_{0}^2\ket \int_{0}^{\tau}j(s)^3\left(i\lambda_{+}e^{\lambda_{+}\left(\tau-s\right)}-i\lambda_{-}e^{\lambda_{-}\left(\tau-s\right)}\right)ds\right.\notag\\
&&+\left.\bra p_{0}^2\ket \int_{0}^{\tau}j(s)g(s)^2\left(i\lambda_{+}e^{\lambda_{+}\left(\tau-s\right)}-i\lambda_{-}e^{\lambda_{-}\left(\tau-s\right)}\right)ds\right.\notag\\
&&+\left.\int_{0}^{\tau}j(s)\bra h(s)^2 \ket \left(i\lambda_{+}e^{\lambda_{+}\left(\tau-s\right)}-i\lambda_{-}e^{\lambda_{-}\left(\tau-s\right)}\right)ds\right]
\end{eqnarray}
The final term is an average over the sum of thermal and radiation pressure noise. As a reminder the former is approximated to be white noise but the latter takes the more complicated form in \eqref{eq:radcorrel}

After the integrals that appear in the above expression can be evaluated and after some straightforward manipulations we arrive at:
\begin{eqnarray}
\label{eq:deltax}
\bra \delta x\left(\tau\right)x_{0}\left(0\right) \ket_{R}&=&\left[-4A\bra x_{0}^2\ket \left(\frac{\lambda_{+}\lambda_{-}}{\left(\lambda_{+}-\lambda_{-}\right)^2}\right)\right]\left\{e^{3\lambda_{+}\tau}\left[\bra x_{0}^2\ket \left(\lambda_{+}\lambda_{-}m\right)^2\left(\frac{-3}{2\left(\lambda_{-}-\lambda_{+}\right)\left(3\lambda_{+}-\lambda_{-}\right)}\right)\right.\right.\notag\\
&&\left.\left.+\bra p_{0}^2\ket \left(\frac{-3\lambda_{+}^2}{2\left(3\lambda_{+}-\lambda_{-}\right)\left(\lambda_{-}-\lambda_{+}\right)}\right)+U\left(\frac{-3\lambda_{+}}{4\left(3\lambda_{+}-\lambda_{-}\right)\left(\lambda_{-}-\lambda_{+}\right)}\right)+E_{1}\right]\right.\notag\\
&&+\left.e^{\left(2\lambda_{+}+\lambda_{-}\right)\tau}\left[\bra x_{0}^2\ket \left(\lambda_{+}\lambda_{-}m\right)^2\left(\frac{3}{2}\left(\frac{2\lambda_{+}+\lambda_{-}}{\lambda_{+}\left(\lambda_{+}+\lambda_{-}\right)\left(\lambda_{-}-\lambda_{+}\right)}\right)\right)\right.\right.\notag\\
&&+\left.\left.\bra p_{0}^2\ket \left(\frac{2\left(\lambda_{+}+\lambda_{-}\right)^2+\lambda_{+}\lambda_{-}}{2\left(\lambda_{+}+\lambda_{-}\right)\left(\lambda_{-}-\lambda_{+}\right)}\right)+U\left(\frac{ \left(2 \lambda _++\lambda _-\right) \left(\lambda _++5 \lambda _-\right)}{4 \left(\lambda _++\lambda _-\right){}^2\left(\lambda_{-}-\lambda_{+}\right)}\right)+E_{2}\right]\right.\notag\\
&&+\left.e^{\lambda_{-}\tau}\left[\bra x_{0}^2\ket \left(\lambda_{+}\lambda_{-}m\right)^2\left(\frac{3}{2}\left(\frac{\lambda_{-}-\lambda_{+}}{\lambda_{+}\left(3\lambda_{+}-\lambda_{-}\right)\left(\lambda_{+}+\lambda_{-}\right)}\right)\right)+\bra p_{0}^2\ket \left(\frac{\lambda_{-}\left(\lambda_{-}-\lambda_{+}\right)}{2\left(\lambda_{+}+\lambda_{-}\right)\left(3\lambda_{+}-\lambda_{-}\right)}\right)\right.\right.\notag\\
&&+\left.\left.U\left(\frac{3 \lambda _+^3+7 \lambda _- \lambda _+^2-3 \lambda _-^2 \lambda _++\lambda _-^3}{2 \left(3 \lambda _+-\lambda _-\right) \left(\lambda _++\lambda _-\right){}^2 \left(\lambda _--\lambda _+\right)}\right)+E_{3}\right.\right]\notag\\
&&-\left.\tau e^{\lambda_{-}\tau}\left(U\left(\frac{\lambda_{-}}{2\left(\lambda_{+}+\lambda_{-}\right)}\right)+E_{4}\right)- e^{(2 \lambda _+ -\frac{\kappa }{2})\tau}E_{5}+e^{\left(\lambda _++\lambda _--\frac{\kappa }{2}\right) \tau }E_{6} +c.c.\right\}
\end{eqnarray}
where $U=2k_B T\gamma m $ and:
\begin{eqnarray}
E_1&=&\frac{3\alpha ^2 G^2 \hbar ^2 \lambda _+ }{\left(\lambda _+-\lambda _-\right) \left(3 \lambda _+-\lambda _-\right) \left(\kappa +2 \lambda _+\right)}\\
E_2&=&-\frac{\alpha ^2 G^2 \hbar ^2\left(2 \lambda _++\lambda _-\right) \left(\lambda _+ \left(\kappa +6 \lambda _-\right)+\lambda _- \left(5 \kappa +6 \lambda _-\right)\right) }{\left(\lambda _+-\lambda _-\right) \left(\lambda _++\lambda _-\right){}^2 \left(\kappa +2 \lambda _+\right) \left(\kappa +2 \lambda _-\right)}\\
E_3&=&[2 \alpha ^2 G^2 \hbar ^2 \left(-3 \kappa ^2 \lambda _+^3 \left(\kappa -2 \lambda _+\right) \left(\kappa -4 \lambda _+\right)+\lambda _-^4 \left(5 \kappa ^3-50 \kappa ^2 \lambda _++44 \kappa  \lambda _+^2-24 \lambda _+^3\right)\right.\notag\\
&&\left.+\lambda _+ \lambda _-^2 \left(\kappa -3 \lambda _+\right) \left(3 \kappa ^3-2 \lambda _+ \left(7 \kappa ^2+8 \lambda _+ \left(2 \kappa +\lambda _+\right)\right)\right)+\kappa  \lambda _+^2 \lambda _- \left(-7 \kappa ^3+47 \kappa ^2 \lambda _+-114 \kappa  \lambda _+^2+72 \lambda _+^3\right)\right.\notag\\
&&\left.-\lambda _-^3 \left(\kappa ^4-\kappa ^3 \lambda _+-82 \kappa ^2 \lambda _+^2+140 \kappa  \lambda _+^3+72 \lambda _+^4\right)+4 \lambda _-^6 \left(\kappa -6 \lambda _+\right)\right.\notag\\
&&\left.+4 \lambda _+ \lambda _-^5 \left(\kappa +18 \lambda _+\right)\right) ][\left(\lambda _--3 \lambda _+\right) \left(\lambda _++\lambda _-\right){}^2 \left(\lambda _--\lambda _+\right) \left(\kappa -2 \lambda _+\right){}^2 \left(\kappa -2 \lambda _-\right){}^2 \left(\kappa -4 \lambda _++2 \lambda _-\right)]^{-1}\\
E_4&=&\frac{2\alpha ^2 G^2 \hbar ^2 \kappa \lambda _- }{\left(\lambda _++\lambda _-\right) \left(\kappa -2 \lambda _-\right) \left(\kappa -2 \lambda _+\right)}\\
E_5&=&\frac{16 \alpha ^2 G^2 \hbar ^2 \kappa  \lambda _+ \left(\kappa -4 \lambda _+\right) }{\left(\kappa -2 \lambda _+\right){}^2 \left(\kappa +2 \lambda _+\right) \left(\kappa +2 \lambda _-\right) \left(\kappa -4 \lambda _++2 \lambda _-\right)}\\
E_6&=&\frac{16 \alpha ^2 G^2 \hbar ^2 \kappa  \lambda _+ \left(\kappa -2 \lambda _+-2 \lambda _-\right) }{\left(\kappa -2 \lambda _+\right){}^2 \left(\kappa +2 \lambda _+\right) \left(\kappa ^2-4 \lambda _-^2\right)}
\end{eqnarray}

Recall that this expression is valid only when $\tau\geq 0$. We can infer the value for negative $\tau$ through the identities $\langle x(\tau)x(0)\rangle^*=\langle x(0)x(\tau)\rangle = \langle x(-\tau)x(0)\rangle$. The first equality follows directly from the definition of the correlator in terms of the steady state density matrix and the time evolution operator. The second equality results from the fact that the system is assumed to be in steady state. Notice that these equalities imply that $\langle x(\tau)x(0)\rangle_R=\langle x(-\tau)x(0)\rangle_R$.

We can compare the terms arising here with the contribution due to the perturbation of the steady state given in equation (\ref{eq:pertcorrelation}). Using thermal expectation values such as $\langle x_0^2\rangle_{H_{0}}=k_BT'/m\Omega^2$ we see from equation (\ref{eq:pertcorrelation}) a perturbation to the correlation function that scales like $Ap_{\rm ZPF}^4/m^2\Omega^2$ in the limit that the oscillator is high-$Q$. Studying the terms in equation (\ref{eq:deltax}) that have frequencies close to $\omega_0$ it is possible to see that in the high-$Q$ regime these contributions scale like $A(k_BT')^2\gamma/\Omega^3$ and other contributions that are smaller by some power of $\gamma/\Omega$. The earlier contributions are therefore different by a factor $(p_{\rm ZPF}^2/mk_BT')^2\Omega/\gamma$. Since $p_{\rm ZPF}^2/mk_BT'$ is the ratio between the momentum variance of the oscillator in the ground state and the thermal state we expect that this term is only relevant very close to zero temperature. However even at high temperatures if the quality factor $\Omega/\gamma$ is sufficiently large the overall term may be significant, and indeed in the tabletop experiments we consider this is the case. Therefore we must keep both contributions in our analysis of the perturbed correlation function.


\subsection{Perturbed spectrum}
Now we can combine the results of the previous sections to obtain the perturbed spectrum. From the expressions for eigenvalues in \eqref{eq:eigenvalues} we can classify the terms in \eqref{eq:deltax} according to their exponents in $\omega_{0}$. 
 The assumption $\kappa\gg\gamma$, typical for optomechanical experiments, allows us to treat the unperturbed oscillator as being in a steady state with an approximate effective temperature $T'$ resulting from the effect of the thermal bath and the radiation pressure fluctuations. The equipartition relations  $\bra p_{0}^2\ket/2m=m\Omega^2\bra x_{0}^2\ket/2=k_{B}T'/2=k_{B}T/2+\hbar ^2\alpha ^2 G^2 \kappa  /2\gamma _0 m \left(\kappa ^2+4 \omega _0^2\right)$ then approximately hold (see section \ref{steady}).  Decomposing the coefficients into real and imaginary parts and adding the quantity in \eqref{eq:pertcorrelation} the perturbed correlation function becomes (for $\tau\geq 0$):
\ba
\delta\bra x_{0}\left(\tau\right)x_{0}\left(0\right) \ket_{R}&\simeq A \bra x_{0}^2\ket \frac{\gamma_{0}^2+\omega_{0}^2}{\omega_{0}^2} \left[e^{\left(-3\gamma_{0}+i3\omega_{0}\right)\tau}W+e^{\left(-\gamma_{0}-i\omega_{0}\right)\tau}\left(I_1+I_2+(J_1+J_2) i\right)-\tau e^{\left(-\gamma_{0}-i\omega_{0}\right)\tau}\left(M+Ni\right)\right.\notag\\
&\left.+e^{(-\frac{\kappa}{2}
	+i2\omega_{0})\tau}(Y+Zi)+e^{(-2\gamma_0-\frac{\kappa}{2})\tau}R+c.c.\right]
\end{align}
where: 
\begin{eqnarray}
W&=&-\frac{3\alpha ^2 G^2 \hbar ^2   \omega _0^2}{ \left(\gamma _0^2+4 \omega _0^2\right) \left(\kappa ^2+4 \omega _0^2\right)}\\
I_1+J_1 i&=&\frac{39\alpha ^2 G^2 \hbar ^2 \omega _0^2  \kappa ^4}{\left(\gamma _0^2+4 \omega _0^2\right) \left(\kappa ^2+4 \omega _0^2\right){}^2 \left(\kappa ^2+36 \omega _0^2\right)}+i \left[\frac{k_{B}mT'\gamma_{0}}{\omega_{0}}-\frac{8\alpha ^2 G^2 \hbar ^2\omega _0\kappa(5\kappa^2+18\omega_{0}^2)(\gamma_0^2+3\omega_0^2)}{ \left(\gamma _0^2+4 \omega _0^2\right) \left(\kappa ^2+4 \omega _0^2\right){}^2 \left(\kappa ^2+36 \omega _0^2\right)}\right]\\
I_2+J_2 i&=&\frac{\hbar^2\omega_{0}^2m}{6k_{B}T'}+i \left[-\frac{\hbar^2\omega_{0}\gamma_{0}m}{6k_{B}T'}\right]\\
M+Ni&=&\gamma _0 k_B m T'+ik_B m T' \omega _0\\
Y+Zi&=&
\frac{16 \alpha ^2 G^2 \hbar ^2\kappa  \left(\gamma _0 \kappa ^4+6 \kappa ^3 \omega _0^2+88 \kappa  \omega _0^4\right)}{\left(\kappa ^2+4 \omega _0^2\right){}^3 \left(\kappa ^2+36 \omega _0^2\right)}+i\left[-\frac{16\alpha ^2 G^2 \hbar ^2 \kappa  \omega _0 \left(\kappa ^4+12 \kappa ^2 \omega _0^2-96 \omega _0^4\right)}{\left(\kappa ^2+4 \omega _0^2\right){}^3 \left(\kappa ^2+36 \omega _0^2\right)}\right]\\
R&=&-\frac{16 \alpha ^2 G^2 \hbar ^2\kappa ^2 \left(\gamma _0 \kappa +2 \omega _0^2\right)}{\left(\kappa ^2+4 \omega _0^2\right){}^3}
\end{eqnarray}
Here $I_1$ and $J_1$ comes from the second term of \eqref{eq:correlation} and $I_2$ and $J_2$ arise from the first term, due to the zero-point energy of the oscillator.

The existence of a time-weighted sinusoidal term in the correlation function is a property of an effectively amplitude-dependent resonance frequency, a hallmark of non-linearity, in a perturbative approximation. In the frequency regime where radiation pressure noise can modelled by white noise
we can take $\kappa\gg \Omega,\gamma$ keeping terms of order $k_{B}T'$. 
In this case we find $W=I_1=Y=Z=R=0$, $J_1=k_{B}mT'\gamma_{0}/\omega_{0}$ and the other terms are the same as before. The perturbed spectrum in this frequency regime is given by:
\begin{eqnarray}
\label{eq:dSfull}
\delta S\left(\omega\right)&=&\int_{-\infty}^{\infty}\delta\bra x_{0}\left(\tau\right)x_{0}\left(0\right)\ket_R e^{i\omega \tau} d\tau\notag\\
&=&\frac{16 A \gamma  k_{B}^2 T'^2 \omega ^2 (\omega^2 -\Omega^2 )}{\left(\gamma ^2 \omega ^2+\left(\omega ^2-\Omega ^2\right)^2\right)^2}+\frac{2A \gamma  \omega ^2 \hbar ^2}{3 \left(\gamma ^2 \omega ^2+\left(\omega ^2-\Omega ^2\right)^2\right)}
\end{eqnarray}

We are mainly interested in the case of high-$Q$ oscillators ($Q=\Omega/\gamma$ for viscous damping) and in specific regimes of frequency in which it is possible to obtain simpler expressions. 
So for example, at resonance $\omega=\Omega$ we obtain:
\ba
\delta S\left(\omega\right)&= \frac{2 A \hbar ^2}{3 \gamma }\hspace{2cm}
\end{align}
Here only the contribution from oscillator quantum zero-point energy remains (up to first order in $A$), and in fact this is the optimal frequency for this term.

Near resonance at $\omega=\Omega\pm\gamma_{0}/\sqrt{3}$ the magnitude of the first term of the perturbed spectrum \eqref{eq:dSfull} is approximately at maximum:
\ba
\delta S\left(\omega\right)\simeq \pm  \frac{A\hbar^2}{\gamma}\left[3\sqrt{3}\left( \frac{k_{B} T'}{\hbar\Omega }\right)^2\frac{\Omega}{\gamma}+\frac{ 1 }{2  }\right]
\end{align}
The approximation sign indicates we have taken the limit of high $Q$.

Another relevant frequency is $\omega=\Omega\pm\gamma_{0}$ where for high $Q$ the ratio between the first term of the perturbed spectrum and the sum of the standard thermal and radiation pressure spectra is maximum. Here:
\ba
\delta S\left(\omega\right)\simeq \pm  \frac{A\hbar^2}{\gamma}\left[4\left( \frac{k_{B} T'}{\hbar\Omega }\right)^2\frac{\Omega}{\gamma}+\frac{ 1 }{3  }\right]
\end{align}

In the free-mass limit $\omega\gg \Omega$ we have:
\ba\label{eq:fulllarge1}
\delta S\left(\omega\right)\simeq \frac{2A\gamma\hbar^2}{\omega^2}\left[8\left(\frac{k_BT'}{\hbar\Omega}\right)^2\left(\frac{\Omega}{\omega}\right)^2+\frac{    1}{3 }\right]
\end{align}
and at low frequencies in the range $\Omega/Q\ll \omega \ll \Omega$:
\ba
\delta S\left(\omega\right)\simeq \frac{2 A\hbar^2 \gamma  \omega ^2 }{\Omega ^4}\left[-8\left(\frac{k_BT'}{\hbar\Omega}\right)^2+\frac{    1}{3 }\right]
\end{align}

Over general frequencies the perturbed spectrum is given by:
\begin{eqnarray}\label{generalspectrum}
\delta S\left(\omega\right)&=&\frac{-4A k_B T'}{m \omega _0^2}\left[
\frac{ \omega ^2 \left((J_1+J_2) \omega _0-\gamma _0 (I_1+I_2)\right)-\left(\gamma _0^2+\omega _0^2\right) \left(\gamma _0 (I_1+I_2)+(J_1+J_2) \omega _0\right)}{2 \omega ^2 \left(\gamma _0-\omega _0\right) \left(\gamma _0+\omega _0\right)+\left(\gamma _0^2+\omega _0^2\right){}^2+\omega ^4}\right.\notag\\
&&\left.+\frac{1 }{\left(2 \omega ^2 \left(\gamma _0-\omega _0\right) \left(\gamma _0+\omega _0\right)+\left(\gamma _0^2+\omega _0^2\right){}^2+\omega ^4\right){}^2}\left[\left(\omega ^4 \left(-\gamma _0^2M+M \omega _0^2-6 \gamma _0 N \omega _0\right)+\omega ^2 \left(M \left(10 \gamma _0^2 \omega _0^2+\gamma _0^4+\omega _0^4\right)\right.\right.\right.\right.\notag\\
&&\left.\left.\left.\left.-4 \gamma _0 N \omega _0 \left(\gamma _0-\omega _0\right) \left(\gamma _0+\omega _0\right)\right)+\left(\gamma _0^2+\omega _0^2\right){}^2 \left(\gamma _0^2 M-M \omega _0^2+2 \gamma _0 N \omega _0\right)-M \omega ^6\right)\right]-\frac{2 R \kappa }{\kappa ^2+4 \omega ^2}\right.\notag\\
&&\left.-\frac{3 W \gamma_0 \left(\omega ^2 +9\gamma _0^2+9 \omega _0^2\right) }{6 \omega ^2 \left(\gamma _0+ \omega _0\right) \left(\gamma _0- \omega _0\right)+81\left(\gamma _0^2+ \omega _0^2\right){}^2+\omega ^4}
+\frac{ \omega ^2 \left(-\frac{Y\kappa }{2}-2 \omega _0 Z\right)-\left(\frac{\kappa^2 }{4}+4\omega _0^2\right) \left(\frac{Y\kappa }{2}-2 \omega _0 Z\right)}{2 \omega ^2 \left(\frac{\kappa }{2}+2 \omega _0\right) \left(\frac{\kappa }{2}-2 \omega _0\right)+\left(\frac{\kappa^2 }{4}+4 \omega _0^2\right){}^2+\omega ^4}\right]
\end{eqnarray}

\subsection{aLIGO modelling}\label{ligomodelling} 
In the main text we analysed the modified commutator signal to get the spectrum. Doing this requires a translation of parameters between the setup of aLIGO involving multiple coupled cavities and mirrors to our setup that has a single cavity, oscillator and optical drive. In this section we explain our approach to obtaining parameters of our model pertaining to aLIGO.

We can model the four identical mirrors in aLIGO as a single mirror in our setup with a reduced mass $m/4$ where $m$ is the mass of each mirror in aLIGO \cite{kimble2001conversion}. This comes from the fact that aLIGO measures the difference between the differences in positions of centre of mass of the mirrors in each arm individually. The $\gamma$ parameter above is commonly modelled in mechanical experiments as a constant over frequency, in which case the damping term in the equation of motion is termed `viscous damping', proportional to the momentum. But for aLIGO the viscous damping of the mirrors, caused by collisions with surrounding gas molecules and control electromagnets, is small as the mirrors sit in an ultra-high vacuum. 

A form of non-viscous damping known as `structural damping' is dominant within the suspension fibres of aLIGO and visible in other experiments with low viscous noise. In the frequency domain the equation of motion is modelled in \cite{saulson1990thermal} via a complex spring constant that is frequency-independent:
\ba\label{structuralfreq}
m\ddot{x}=-k(1+i\phi)(x-x_{g})+F
\end{align}
where $x/x_{g}=\Omega^2(1+\phi)/(\Omega^2-\omega^2+i\phi\Omega^2)$ is the vibration transfer function from motion at a point $x_g$ to $x$.
This frequency-independence is a good approximation at frequencies away from zero for realistic structural damping (since close to zero frequency the true loss must drop to zero). The equation of motion can be expressed in the time domain via a Hamiltonian as follows so that its associated equation of motion is equivalent to \eqref{structuralfreq}.

 Consider a mirror system oscillator that is in contact with a bath modelled as consisting of many oscillators with weak coupling to the oscillator. Solving the Heisenberg equations of bath and oscillator simultaneously, using the entire system and bath Hamiltonian, we obtain the following equation of motion for the oscillator \cite{gardiner2004quantum} (with initial time $t_{0}=-\infty$):
\ba\label{bath}
m\ddot{x}=-k(x-x_g)+F_\text{modified}-\int_{-\infty}^{t}\frac{d(x-x_g)}{dt'}f(t'-t)dt'+F_\text{noise}
\end{align}
Here $F_\text{noise}$ is the total fluctuating noise force and $F_\text{modified}$ is the effective force due to the modification of the canonical commutator. $F_\text{noise}$ possesses approximately Gaussian statistics of an equilibrium thermal ensemble with a symmetrised time correlation function proportional to $f(t'-t)$. The third term in \eqref{bath} is the effective dissipation force on the system. For viscous damping $f(t'-t)=2\gamma m \delta(t'-t)$. For structural damping we set $f(t'-t)=-2\gamma m\Omega(\gamma_{EM}+\ln\abs{t'-t}))$, where $\gamma_{EM}$ is the Euler-Mascheroni constant, to match the frequency domain expression \eqref{structuralfreq} as in \cite{saulson1990thermal} associated with frequency-independent loss factor $\phi$.

 Now in the regime of high frequencies, which is of interest in the free-mass limit, structural damping deviates significantly from viscous damping. Here we present a heuristic argument that in this regime structural damping may be viewed as a frequency dependent version of viscous damping. The dominant contribution to the spectrum at high frequencies is from motion in which the interval of time  $\abs{t'-t}$ in \eqref{bath} is small so here we may make the approximation $d(x-x_g)/dt'\sim d(x-x_g)/dt$ and move it out of the integral.  We can impose cutoffs of the remaining integral as we care about high frequencies/small times. We choose the cutoffs such that we recover from the equation of motion, in the case of standard canonical commutator, the position spectrum associated with structural damping as defined in \cite{saulson1990thermal}. The remaining integral then equals $\gamma m\left(\Omega/\omega\right)$. We may thus interpret the dissipation term as involving viscous damping but with an effectively renormalised $\gamma$ that is now frequency dependent: $\gamma'=\Omega/Q\left(\Omega/\omega\right)$. By using $\gamma'$ instead of $\gamma$ throughout our derivation of the perturbed spectrum we may arrive at the approximate perturbed spectrum with structural damping.

We also have to translate the optical parameters in aLIGO, which we can do by matching the radiation pressure and shot noise spectra of our setup to that of aLIGO. In our setup the standard radiation pressure noise spectrum is $4\hbar^2 \alpha ^2 G^2/\kappa\left(4 \omega ^2/\kappa ^2+1\right)   m^2 \left(\gamma ^2 \omega ^2+\left(\omega ^2-\Omega ^2\right)^2\right)\simeq 16 h\nu P \mathcal{F}^2/\pi^2 c^2m^2\left(4 \omega ^2/\kappa ^2+1\right)\omega^4$ where the latter is valid for $\omega\gg\Omega\gg\gamma$. In aLIGO it is in the same frequency regime given by  $ 64h \nu G_{-}P_{arm}\left(\left(2\pi f_{-}\right)^2/\omega^2+1\right)/c^2M^2\omega^2$ where $G_{-}=31.4$ is the cavity build-up factor for the differential mode in aLIGO (difference between positions of mirrors in each arm; the degree of freedom aLIGO measures) and $f_{-}$ is the differential coupled cavity pole \cite{martynov2016sensitivity}. The laser frequency in our setup is taken to be the same as used in aLIGO. We now take $m=M/4$ as explained above, and assuming the power in each cavity of aLIGO's arms and our setup are the same then the cavity build-up factor in our setup is $P_{arm}/P=2\mathcal{F}/\pi$ where $P_{arm}$ is the power in each cavity of aLIGO. The spectra are then equal to each other provided that $\mathcal{F}=\pi G_{-}/2$ which implies $P=P_{arm}/G_{-}$, and, keeping the same arm-length $L$ as in aLIGO, that the optical decay rate is $\kappa=4\pi f_{-}=2 c/G_{-}L$. 

To translate the shot noise spectrum we first have that in our setup $\kappa/16 \alpha ^2 G^2\eta_2\left(4 \omega ^2/\kappa ^2+1\right)=h\nu\lambda^2/256P\mathcal{F}^2\eta_2\left(4 \omega ^2/\kappa ^2+1\right)$. Here we have introduced a parameter $\eta_2$, effectively a detection efficiency factor, in order that our shot noise spectrum matches that in aLIGO. In aLIGO the shot noise spectrum is given by $2h\nu\lambda^2G_{src}/G_{prc}P_{in}\eta G_{arm}^2\left(4\pi\right)^2\left(\left(2\pi f_{-}\right)^2/\omega^2+1\right)$ where $P_{in}$ is the input power to the interferometer, $G_{prc}$ is the power recycling cavity gain, $\eta=0.75$ is the detection inefficiency of aLIGO (fraction of the output power that is transmitted to the photodiode detectors), $G_{arm}$ is the buildup factor of each arm cavity of aLIGO and $G_{src}$ is the signal recycling cavity gain of aLIGO (these are expressed in rounded off form in \cite{martynov2016sensitivity}). Substituting $\mathcal{F}=\pi G_{-}/2$, $P=P_{in}G_{arm}G_{prc}/2G_{-}$ which equates power in our setup cavity to that in aLIGO's arm cavities, and $G_{-}=G_{arm}/G_{src}$  we arrive at $\eta_2=\eta/4$. The relevant translated parameters are contained in table \ref{table:table2}.

\subsection{Steady-state expectation values without the adiabatic approximation}\label{steady}
Here we sketch the calculation of the steady-state expectation values of position/momentum variances, if the adiabatic approximation for the cavity field is not made.

 Recall that the radiation pressure force driving of the cavity is $f_{\rm rad}=\hbar G \alpha \left(\delta a +\delta a^\dagger\right)$. If the cavity is driven on resonance the required quadrature of the cavity field can be used to show that $f_{\rm rad}$ satisfies the equation \cite{aspelmeyer2014cavity}
\begin{equation}
\dot{f}_{\rm rad}= -\kappa f_{\rm rad}/2 +\sqrt{\kappa} \hbar G\alpha \left(f_{\rm cav}+f_{\rm cav}^\dagger\right),
\end{equation}
where $f_{\rm cav}$ is the vacuum field input to the cavity and can be modelled by white noise with $\langle f_{\rm cav}(t)f_{\rm cav}^\dagger (t')\rangle =\delta(t-t') $. Notice that this equation holds even for non-zero $A$. Consequently we have
\begin{equation}
f_{\rm rad}(t) = e^{-\kappa t/2}f_{\rm rad}(0)+ \sqrt{\kappa}\hbar G\alpha\int_0^t e^{-\kappa (t-t')/2}\left[f_{\rm cav}(t')+f_{\rm cav}^\dagger(t')\right]dt'.
\end{equation}

Consequently in the steady state we have
\begin{equation}
\langle f_{\rm rad}(t)f_{\rm rad}(t')\rangle = \hbar^2 G^2\alpha^2e^{-\kappa |t-t'|/2}.
\end{equation}
When the assumption is made that the oscillator dynamics are slow compared to $\kappa$ this exact correlator can be approximated well by the appropriate delta-function.

We also need to determine the contribution of the radiation pressure force to $\langle x_0^2\rangle$ and the other steady state expectation values of the unperturbed system. These can easily be found using equation (\ref{eq:solutions1}). So we have
\begin{eqnarray}
\langle x_0^2\rangle &=& \lim_{t\rightarrow \infty} \langle x_0(t)^2\rangle = \frac{1}{4m^2\omega_0^2} \lim_{t\rightarrow \infty} \langle (ia-ia^*)^2
\rangle \notag \\
&=&\frac{\hbar ^2\alpha ^2 G^2 (\kappa+4\gamma_0)}{\gamma _0 m^2\Omega^2 \left((\kappa+2\gamma_0) ^2+4 \omega _0^2\right)}
\end{eqnarray}
On the other hand
\begin{eqnarray}
\langle p_0^2\rangle &=& \lim_{t\rightarrow \infty} \langle p_0(t)^2\rangle = \frac{1}{4\omega_0^2} \lim_{t\rightarrow \infty} \langle (-i\lambda_-a+i\lambda_+a^*)^2
\rangle \notag \\
&=&\frac{\hbar ^2\alpha ^2 G^2 \kappa  }{\gamma _0  \left((\kappa+2\gamma_0) ^2+4 \omega _0^2\right)}
\end{eqnarray}

\subsection{Standard spectrum}\label{standardspectrum}
We remind the reader that for the standard canonical commutator the total noise spectrum, modelled as the sum of thermal, radiation pressure noise and shot noise spectra that we have used above, is given by \cite{aspelmeyer2014cavity}\cite{clerk2010introduction}:
\ba\label{standardspec}
S_\text{std}(\omega)=\frac{2 \gamma  k T}{m \left(\gamma ^2 \omega ^2+\left(\omega ^2-\Omega ^2\right)^2\right)}+\frac{4\hbar^2 \alpha ^2 G^2}{\kappa\left(\frac{4 \omega ^2}{\kappa ^2}+1\right)   m^2 \left(\gamma ^2 \omega ^2+\left(\omega ^2-\Omega ^2\right)^2\right)}+\frac{\kappa  \left(\frac{4 \omega ^2}{\kappa ^2}+1\right)}{16 \alpha ^2 G^2}
\end{align}

\bibliography{modcomm}

\end{document}